**ORIGINAL PAPER**

# An efficient algorithm for rigid/deformable contact interaction based on the dual mortar method

R. Pinto Carvalho[1,2] · A. M. Couto Carneiro[1,2] · F. M. Andrade Pires[1,2] · A. Popp[3]

## Abstract

In a wide range of practical problems, such as forming operations and impact tests, treating one of the contacting bodies as a rigid obstacle is an excellent approximation to the physical phenomenon. In this work, the well-established dual mortar method is adopted to enforce interface constraints in the finite deformation frictionless contact of deformable bodies and rigid obstacles. The efficiency of the non-linear contact algorithm proposed here is based on two main contributions. Firstly, the weighted gap function is modified such that it retains the signal of the discrete gap function. Within the context of rigid/deformable contact, this unlocks a significant simplification by removing the need to explicitly evaluate the dual basis functions. The corresponding first-order interpolation is presented in detail. Particular focus is, then, placed on the extension for second-order interpolation by employing a piecewise linear interpolation scheme, which critically retains the geometrical information of the finite element mesh. Secondly, a new definition for the nodal orthonormal moving frame attached to each contact node is suggested. It reduces the geometrical coupling between the nodes and consequently decreases the stiffness matrix bandwidth. The proposed contributions decrease the computational complexity of dual mortar methods for rigid/deformable interaction, especially in the three-dimensional setting, while preserving accuracy and robustness.

## 1 Introduction

The application of the Finite Element Method (FEM) to contact problems can be traced back to the late 1970s and early 1980s, with contributions such as [1–5] being the initial spark for what would become the field of computational contact mechanics. The classical discretisation techniques were based on purely nodal considerations, with the node-to-segment algorithm being, arguably, the most widespread technique since then. Over the last decades, however, alternative formulations for improved robustness have gained considerable attention, particularly the application of the so-called mortar methods for contact problems. Essentially, the concept is based on the establishment of a variational operator for the imposition of occurring interface constraints in an optimal weak sense. In the context of contact problems, it allows the treatment of the contact constraints for non-matching interface discretisations typically found in finite deformation settings. Early implementations of mortar methods within the context of small deformation contact problems can be found in [6–8]. Subsequently, the extension into the realm of fully non-linear large deformation kinematics was gradually derived and, without claiming the following list to be exhaustive, the reader is referred to [9–16].

A fundamental aspect of mortar methods is that the solution is typically enforced by using Lagrange multipliers, which, in order to preserve the accuracy of the solution, need to be judiciously chosen. Nowadays, a popular choice for the Lagrange multiplier space is the so-called dual Lagrange multiplier approach [17]. In contrast with the standard choice of mortar methods, dual Lagrange multipliers are of particular interest because they generate coupling conditions that are easier to realise, without compromising on the optimal convergence of the discretisation error. This fundamental

✉ A. Popp
alexander.popp@unibw.de

1 Department of Mechanical Engineering, University of Porto, Rua Dr. Roberto Frias, 4200-465 Porto, Portugal

2 Institute of Science and Innovation in Mechanical and Industrial Engineering (INEGI), Rua Dr. Roberto Frias, 4200-465 Porto, Portugal

3 Institute for Mathematics and Computer-Based Simulation, Universität der Bundeswehr München, Werner-Heisenberg-Weg 39, 85577 Neubiberg, Germany advantage is deeply associated with the possibility to algebraically condense the Lagrange multipliers from the system of equations in a trivial fashion. Early applications of this technique to small deformation contact problems can be found in [18–21]. Exemplary works focused on the extension to general finite deformation contact problems include [22–26].

A summarized overview of the dual mortar contact algorithm would describe it as a highly versatile and accurate method requiring minimal calibration by the user. However, the trade-off regards its complexity, both in terms of formulation and overall computational implementation and computation. This becomes aggravated when considering quadratic elements, which, for industrial applications, are particularly attractive to model complex geometries and avoid numerical artefacts commonly found in linear elements, such as shear locking, volumetric locking and hourglassing. This renders the motivation for optimisation strategies that aim at reducing the complexity of the contact algorithm, yet without compromising on its mathematical properties.

The classification of contact and impact procedures is typically based on the problem configuration, with factors like the total number of bodies contacting each other or their physical behaviour originating different contact problems. While the contact between two deformable bodies is typically termed as unilateral contact, the setup consisting of one single deformable body going against a rigid obstacle is commonly referred to as Signorini contact. The former can be interpreted as the most general class, turning out essential to derive the fundamental mathematical and computational frameworks. However, the underlying assumptions in Signorini contact promote opportunities for simplifications, and the resulting strategy is an excellent approximation to a wide range of engineering systems, such as forming and impact tests. This work is precisely dedicated to the development of techniques that exploit the simplifications of rigid/deformable contact, having in mind the goal of reducing the computational complexity and improving the robustness of the contact algorithm.

The dual mortar method was first applied to finite deformation contact problems involving rigid boundaries in [18]. There, the assumption of a rigid boundary comes up as a simplification for the formulation, which is not thoroughly explored. Later on, in [27], the Signorini contact problem is also mentioned, although within the context of thin-walled structures. The issue was revisited more recently in [28] to model rough surface contact. A mortar contact formulation for second-order elements—which consists of using different interpolation bases for the Lagrange multipliers and their variation—is proposed, to avoid possible consistency errors of the dual mortar algorithm in the case of large curvatures. In this work, this technique will be further investigated. In particular, the potential within rigid/deformable contact to simplify the algorithm significantly by removing the need to actually evaluate the dual basis functions is demontrated. Moreover, an alternative technique for the interpolation of the variation of the Lagrange multipliers is proposed, which is based on the concept of a piecewise linear interpolation scheme. Compared with the solution proposed in [28], the main advantage is that it retains all the geometrical information of the finite element mesh by allocating Lagrange multipliers to every node in the non-mortar boundary. Moreover, the fact that one of the boundaries remains fixed is exploited and a new definition for the unit normal vector attached to each non-mortar node is proposed. It involves projecting the unit normal from the rigid side to the deformable boundary. Eventually, this becomes an efficient alternative to the averaged unit normal typically employed together with mortar methods. As will be shown in great detail in this work, the computational advantages of the proposed nodal moving frame are deeply connected with the simpler linearisation procedure, which reduces the geometrical coupling between contact nodes.

The remainder of this paper is organised as follows. In Sect. 2, the continuum mechanics relations for rigid/deformable contact under large deformations are introduced, both in strong and weak form. Then, the finite element discretisation is described in Sect. 3, emphasizing the mortar coupling terms and discrete contact constraints. The first central concept of this work is described in Sect. 4 by discussing the modification of the weighted gap. Secondly, a new definition for the nodal orthonormal moving frame attached to each contact node is presented in Sect. 5. The numerical evaluation of mortar integrals is described in Sect. 6, followed by the description of the global solution algorithm in Sect. 7, which includes the consistent linearisation for the application of the semi-smooth Newton algorithm and a suitable algebraic representation of the final condensed system of equations. Then, several numerical examples are presented and discussed in Sect. 8, to validate and inspect the computational performance of the algorithm. Lastly, the main conclusions of this work are drawn in Sect. 9.

## 2 Continuum mechanics of the Signorini contact problem

From the viewpoint of mathematical problem formulation, the Signorini contact problem is a popular alternative to introduce the fundamental concepts of contact mechanics within the linear regime. Nonetheless, the approach to rigid/deformable contact derived in this work assumes large deformations and generic constitutive behaviour. This leads to a fully non-linear problem, which can be obtained based upon the concepts of unilateral contact between two deformable bodies. The problem is derived in its strong and weak form in the following.

## 2.1 Strong formulation

The nomenclature adopted and the representation of the problem are schematically illustrated in Fig. 1. The (only) deformable body is designated as non-mortar and identified with the superscript $(\bullet)^s$. The open set $\Omega_0^s \subset \mathbb{R}^d (d = 2, 3)$ represents its reference configuration and the boundary $\Gamma_c^r$ stands for the rigid obstacle contour. Herein, the superscript $(\bullet)^r$ relates to quantities associated with the rigid boundary (thus replacing the notion of the mortar side, typically found in unilateral contact). The boundary $\partial \Omega_0^s$ in the reference configuration can be divided into three open disjoint subsets: the Dirichlet boundary $\Gamma_u^s$, with prescribed displacements $\bar{u}^s$, the Neumann boundary $\Gamma_\sigma^s$, satisfying a given surface traction $\bar{t}^s$, and the non-mortar potential contact interface $\Gamma_c^s$. As the body undergoes motion, denoted by the smooth mapping $\varphi^s$, the counterparts in the current configuration $\Omega_t^s \subset \mathbb{R}^d$ are referred to as $\gamma_u^s$, $\gamma_\sigma^s$ and $\gamma_c^s$, respectively. Note that the rigid boundary $\Gamma_c^r$ remains stationary throughout the entire process, thus having a null displacement field.

The boundary value problem of finite deformation quasi-statics requires the displacement field, $u^s = x^s - X^s$, which describes the motion between the reference configuration $X^s$ and the current configuration $x^s$, to satisfy the momentum balance principle and the set of Dirichlet and Neumann boundary conditions:

$$\text{div } \sigma^s + b^s = 0, \quad \text{in } \Omega_t^s \times (0, T), \tag{1}$$

$$u^s = \bar{u}^s, \quad \text{on } \gamma_u^s \times (0, T), \tag{2}$$

$$\sigma^s n^s = \bar{t}^s, \quad \text{on } \gamma_\sigma^s \times (0, T). \tag{3}$$

Herein, $t \in [0, T]$ plays the role of a pseudo-time, $\sigma^s$ stands for the Cauchy stress tensor and $b^s$ represents the body force per deformed unit volume. The prescribed displacements and surface tractions at the Dirichlet and Neumann boundaries in the current configuration are represented by $\bar{u}^s$ and $\bar{t}^s$, respectively. Moreover, $n^s$ denotes the outward normal vector to the surface $\gamma_\sigma^s$. The contact conditions lead to an additional set of constraints and are described in the following.

The classical continuum mechanics framework of unilateral contact is equally applicable to this problem, with the only exception being a slight modification in the gap vector definition. As a fundamental entity of contact kinematics, the gap vector, $g(x^s, t)$, is involved in the definition of the gap function,

$$g(x^s, t) = \eta(x^s, t) \cdot g(x^s, t), \tag{4}$$

and for rigid/deformable contact, it follows as

$$g(x^s, t) \equiv x^s - \hat{x}^r(x^s, t). \tag{5}$$

The contact point $\hat{x}^r$ on the rigid boundary $\Gamma_c^r$ stands for the projection of the non-mortar point $x^s \in \gamma_c^s$ along its current outward unit normal vector $\eta$. Compared with the general unilateral contact formulation, the difference is that the coordinates on the opposing side to the non-mortar boundary $\gamma_c^s$ remain fixed, i.e.,

$$x^r \equiv X^r. \tag{6}$$

Nonetheless, the projection point $\hat{x}^r$ itself still depends on the deformation of the opposite contact boundary $\gamma_c^s$ since the projection possibly changes over time. As explained in more detail in the following paragraphs describing the discrete version of the problem, the fact that one of its terms remains fixed simplifies the evaluation of several terms.

Besides the kinematical description of contact, the establishment of contact constraints requires a compatible mathematical description of the forces that develop within the active contact region. Thus, the contact traction $t_c^s(x^s, t)$ acting on the current non-mortar contact region $\gamma_c^s$ is introduced and its decomposition into normal and tangential components yields

$$t_c^s(x^s, t) = p^\eta \eta + t^\tau. \tag{7}$$

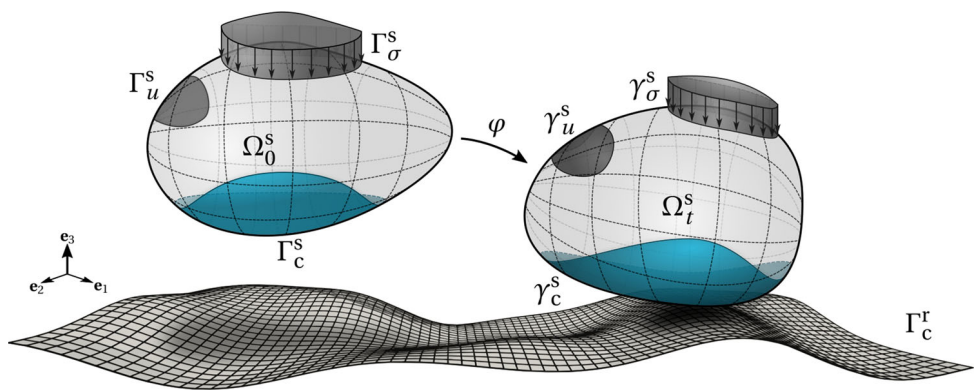

**Fig. 1** Illustration and nomenclature of the Signorini contact problem in the reference and current configurations

The term $p^\eta(\boldsymbol{x}^s, t)$ represents the contact pressure,

$$p^\eta(\boldsymbol{x}^s, t) \equiv \boldsymbol{\eta} \cdot \boldsymbol{t}_c^s, \tag{8}$$

and the frictional traction vector, $\boldsymbol{t}^\tau(\boldsymbol{x}^s, t)$, results from the projection on the tangential plane, i.e.,

$$\boldsymbol{t}^\tau(\boldsymbol{x}^s, t) \equiv (\boldsymbol{I}_d - \boldsymbol{\eta} \otimes \boldsymbol{\eta}) \boldsymbol{t}_c^s. \tag{9}$$

Here, $\boldsymbol{I}_d$ stands for the $d$-dimensional identity tensor. Altogether, the classical Hertz-Signorini-Moreau (HSM) conditions along the normal direction are given by

$$g \geq 0, \quad p^\eta \leq 0, \quad gp^\eta = 0, \quad \text{on } \gamma_c^s. \tag{10}$$

The frictionless tangential condition reads

$$\boldsymbol{t}^\tau = \boldsymbol{0}, \quad \text{on } \gamma_c^s. \tag{11}$$

### 2.2 Weak formulation

The derivation of the weak version of the problem begins with the introduction of the solution space $\mathcal{U}^s$ and weighting space $\mathcal{V}^s$:

$$\mathcal{U}^s \equiv \left\{ \boldsymbol{u}^s \in [H^1(\Omega_t^s)]^d \mid \boldsymbol{u}^s = \bar{\boldsymbol{u}}^s \text{ on } \gamma_u^s \right\}, \tag{12}$$

$$\mathcal{V}^s \equiv \left\{ \delta\boldsymbol{u}^s \in [H^1(\Omega_t^s)]^d \mid \delta\boldsymbol{u}^s = \boldsymbol{0} \text{ on } \gamma_u^s \right\}. \tag{13}$$

These spaces are conceptually similar to the general case of unilateral contact, although involving only the mappings on the deformable non-mortar body. Consideration of the principle of virtual work allows rewriting the momentum balance, Eq. (1), as

$$\delta\Pi_{\text{int}}^s - \delta\Pi_{\text{ext}}^s - \delta\Pi_c^s = 0, \quad \forall\, \delta\boldsymbol{u}^s \in \mathcal{V}^s, \tag{14}$$

where $\delta\Pi_{\text{int}}^s(\boldsymbol{u}^s, \delta\boldsymbol{u}^s)$ represents the internal virtual work,

$$\delta\Pi_{\text{int}}^s(\boldsymbol{u}, \delta\boldsymbol{u}) \equiv \int_{\Omega_t^s} \boldsymbol{\sigma}^s(\boldsymbol{u}^s) : \nabla_x(\delta\boldsymbol{u}^s) \, d\Omega_t^s, \tag{15}$$

and $\delta\Pi_{\text{ext}}^s(\delta\boldsymbol{u}^s)$ the external virtual work,

$$\delta\Pi_{\text{ext}}^s(\delta\boldsymbol{u}) \equiv \int_{\Omega_t^s} \boldsymbol{b}^s \cdot \delta\boldsymbol{u}^s \, d\Omega_t^s + \int_{\gamma_\sigma^s} \bar{\boldsymbol{t}}^s \cdot \delta\boldsymbol{u}^s \, d\gamma_\sigma^s. \tag{16}$$

The third term, $\delta\Pi_c^s(\boldsymbol{u}^s, \delta\boldsymbol{u}^s)$, stands for the virtual work of contact forces. These terms remain unchanged from the classical unilateral contact problem, with the exception being the virtual work of contact forces. For the particular case of Signorini contact, this contribution involves only the virtual displacements of the non-mortar side, i.e.,

$$\delta\Pi_c^s(\boldsymbol{u}^s, \delta\boldsymbol{u}^s) \equiv \int_{\gamma_c^s} \boldsymbol{t}_c^s \cdot \delta\boldsymbol{u}^s \, d\gamma_c^s. \tag{17}$$

As discussed in more detail in Sect. 3.1, this is the origin of one of the main simplifications of the algorithm. The so-called second mortar coupling matrix can be removed, thus evaluating the virtual work of contact forces using solely the first mortar coupling matrix.

The last step towards the formulation of the weak version of rigid/deformable large deformation contact comprises the treatment of the contact constraints. The formulation proposed in this work is based on the dual mortar algorithm, which introduces the Lagrange multiplier as the negative contact traction on the non-mortar side, i.e., $\boldsymbol{\lambda} = -\boldsymbol{t}_c^s$. This sets the basis for a mixed variational approach. Just like the contact traction, the Lagrange multiplier is decomposed into normal and tangential parts, i.e.,

$$\boldsymbol{\lambda} = \lambda^\eta \boldsymbol{\eta} + \boldsymbol{\lambda}^\tau. \tag{18}$$

Based on these considerations, according to [29] the Lagrange multiplier vector is chosen from the convex cone $\mathcal{M}^+ \subset \mathcal{M}$ given by

$$\mathcal{M}^+ \equiv \Big\{ \delta\boldsymbol{\lambda} \in \mathcal{M} \mid \delta\boldsymbol{\lambda}^\tau = \boldsymbol{0}$$
$$\langle \delta\lambda^\eta, w \rangle_{\gamma_c^s} \geq 0, \ w \in \mathcal{W}^s, w \geq 0 \Big\}, \tag{19}$$

where $\langle \bullet, \bullet \rangle_{\gamma_c^s}$ stands for the $H^{1/2}$-duality pairing operator. The term $\mathcal{W}^s$ represents the trace space, i.e. the restriction of the solution space $\mathcal{U}^s$ to the non-mortar contact boundary $\gamma_c^s$. Its dual counterpart on $\gamma_c^s$ is represented by $\mathcal{M}$. Lastly, $\delta\boldsymbol{\lambda}$ represents a trial contact traction. It should be mentioned that the solution cone for the Lagrange multipliers space in Eq. (19) simultaneously satisfies, in a weak sense, the HSM conditions in the normal direction and the frictionless tangential sliding condition.

Lastly, by adopting the concept of variational inequalities, the final weak form of the problem can be stated. Given the internal forces and boundary conditions, the problem consists of finding the kinematically admissible displacement function, $\boldsymbol{u} \in \mathcal{U}^s$, and Lagrange multiplier, $\boldsymbol{\lambda} \in \mathcal{M}^+$, such that, for all $t \in [0, T]$, the virtual work equation

$$\delta\Pi_{\text{int}}^s(\boldsymbol{u}, \delta\boldsymbol{u}) - \delta\Pi_{\text{ext}}^s(\delta\boldsymbol{u}) + \int_{\gamma_c^s} \boldsymbol{\lambda} \cdot \delta\boldsymbol{u}^s \, d\gamma_c^s = 0,$$
$$\forall\, \delta\boldsymbol{u} \in \mathcal{V}^s, \tag{20}$$

the normal contact constraints

$$\langle g, \delta\lambda - \lambda \rangle_{\gamma_c^s} \geq 0, \quad \forall \, \delta\lambda \in \mathcal{M}^+, \tag{21}$$

and frictionless condition

$$\lambda^\tau = \mathbf{0}, \tag{22}$$

are satisfied for any admissible test functions $\delta u \in \mathcal{V}^s$ and $\delta\lambda \in \mathcal{M}^+$.

## 3 Mortar finite element discretisation

The weak form derived in Sect. 2.2 is approximated using the finite element method. Its foundation relies on the partition of the domain $\Omega^s$ into $n^e$ element subdomains and on the approximation of the geometry and displacement field at discrete points of interest. Mathematically speaking, this introduces the finite-dimensional subsets $\{\mathcal{U}^s\}^h \subset \mathcal{U}^s$ and $\{\mathcal{V}^s\}^h \subset \mathcal{V}^s$ as approximations for their corresponding functional sets. Following the isoparametric approach, both geometry and field variables are approximated using the same interpolation functions and, additionally, the element domain is mapped to the parameter space $\boldsymbol{\xi}^s = (\xi_1^s, \ldots, \xi_d^s)$. With the focus being on the finite element discretisation of the contact terms, though, only the associated physical quantities are described in the following.

The finite element interpolation at the contact interface reads:

$$\boldsymbol{x}^s \approx \{\boldsymbol{x}^s\}^h \Big|_{\{\gamma_c^s\}^h} \equiv \sum_{k=1}^{n^s} N_k^s(\boldsymbol{\xi}^s) \, \mathbf{x}_k^s, \tag{23}$$

$$\boldsymbol{u}^s \approx \{\boldsymbol{u}^s\}^h \Big|_{\{\gamma_c^s\}^h} \equiv \sum_{k=1}^{n^s} N_k^s(\boldsymbol{\xi}^s) \, \mathbf{d}_k^s. \tag{24}$$

Here, $n^s$ denotes the total number of nodes on the discrete non-mortar boundary $\{\gamma_c^s\}^h$. The corresponding discrete nodal coordinates (in the current configuration) are represented by $\mathbf{x}_k^s$ and the nodal displacement by $\mathbf{d}_k^s$. The shape functions $N_k^s$ are defined with respect to the associated finite element parameter space $\boldsymbol{\xi}^s$.

One important notion to introduce at this stage is that the geometry of the rigid boundary $\Gamma_c^r$ is also discretised, i.e.,

$$\boldsymbol{x}^r \approx \{\boldsymbol{x}^r\}^h \equiv \sum_{l=1}^{n^r} N_l^r(\boldsymbol{\xi}^r) \, \mathbf{x}_l^r, \tag{25}$$

with $n^r$ denoting the total number of nodes defining the rigid boundary. From a conceptual point of view, this step is not mandatory, as the rigid boundary could be equally defined by some analytical function. From a practical perspective, however, finite element interpolation is possibly more convenient, as it allows for the treatment of arbitrarily complex geometries found in engineering applications (which can be difficult to describe analytically). It also allows reusing several procedures and algorithms already established for unilateral contact, such as contact search and the numerical evaluation of mortar integrals. Notwithstanding, it should be kept in mind that all the techniques described in the following are still applicable for analytical representations of rigid boundaries (and, in some cases, even simplified).

To complete the discretisation framework of the problem, the interpolation method of the Lagrange multipliers $\boldsymbol{\lambda}$ must be defined. Within this work, the Lagrange multiplier interpolation is realized on the non-mortar side, and its approximation is based on the introduction of the discrete Lagrange multiplier space $\mathcal{M}^h \subset \mathcal{M}$. Additional details concerning the choice of this discrete space, its fundamental properties and implementation strategy follow in Sect. 4. At this stage, though, a generic notation can be introduced as

$$\boldsymbol{\lambda} \approx \boldsymbol{\lambda}^h \equiv \sum_{j=1}^{n^\lambda} \Phi_j(\boldsymbol{\xi}^s) \, \mathbf{z}_j, \tag{26}$$

with $\Phi_j$ representing the discrete Lagrange multiplier basis functions, $n^\lambda$ the total number of non-mortar nodes carrying additional Lagrange multiplier degrees of freedom and $\mathbf{z}_j$ the discrete nodal Lagrange multipliers. In mortar methods, it is common to consider that every non-mortar node serves as a coupling node, thus $n^\lambda = n^s$. However, for the sake of generality, this is not considered at this stage.

### 3.1 Discrete contact virtual work

Based upon the finite element interpolation scheme introduced in the previous section, the discretised version of the contact virtual work $\delta\Pi_c^s$, given in Eq. (17), can be written as

$$\delta\Pi_c^s \approx \{\delta\Pi_c^s\}^h \equiv \sum_{j=1}^{n^s} \sum_{k=1}^{n^s} \left\{ \mathbf{z}_j \left[ \int_{\{\gamma_c^s\}^h} \Phi_j(\boldsymbol{\xi}^s) N_k^s(\boldsymbol{\xi}^s) \, \mathrm{d}\gamma_c^s \right] \delta\mathbf{d}_k^s \right\}. \tag{27}$$

This introduces a fundamental entity within mortar methods: the first mortar coupling matrix, herein denoted by $[\mathbf{D}] \in \mathbb{R}^{(d \cdot n^\lambda) \times (d \cdot n^s)}$. Its nodal block $\mathbf{D}_{[j,k]}$ is defined as

$$\mathbf{D}_{[j,k]} \equiv \mathbf{I}_d \int_{\{\gamma_c^s\}^h} \Phi_j(\boldsymbol{\xi}^s) N_k^s(\boldsymbol{\xi}^s) \, \mathrm{d}\gamma_c^s, \quad j = 1, \ldots, n^\lambda, \ k = 1, \ldots, n^s. \tag{28}$$

It should be noted that the contact virtual work is entirely based on the first mortar coupling matrix. In contrast to the mortar boundary in unilateral contact problems, the rigid side remains completely prescribed (fixed), it does not contribute to the virtual work by the contact forces—the second mortar matrix is not required.

### 3.2 Discrete contact constraints

The treatment of the contact constraints starts with the definition of the discrete version of the gap vector, $\boldsymbol{g}^h(\boldsymbol{\xi}^s)$, which for rigid/deformable contact reads

$$\boldsymbol{g} \approx \boldsymbol{g}^h(\boldsymbol{\xi}^s) \equiv \sum_{j=1}^{n^s} N_j^s(\boldsymbol{\xi}^s)\mathbf{x}_j^s - \sum_{l=1}^{n^r} N_l^r(\hat{\boldsymbol{\xi}}^r)\mathbf{x}_l^r. \quad (29)$$

When compared with the general case of unilateral contact, they both have the same structure. However, because the derivative of the rigid surface coordinates vanishes, its linearisation is simplified.

As a final remark, it is worth mentioning that if the rigid boundary geometry is discretised using finite elements, the projection operations can be performed by employing existing techniques for unilateral contact, e.g., see [11,12,30]. These typically include an efficient global search algorithm and a continuous field of normals associated with the non-mortar boundary, as explained in mode detail in Sect. 5.

The discrete version of the normal contact constraints, see Eq. (21), leads to the condition

$$\int_{\gamma_c^s} g\left(\delta\lambda^\eta - \lambda^\eta\right) d\gamma_c^s$$
$$\approx \sum_{j=1}^{n^\lambda} \left(\delta z_j^\eta - z_j^\eta\right) \int_{\{\gamma_c^s\}^h} \Phi_j(\boldsymbol{\xi}^s) \, g^h(\boldsymbol{\xi}^s) \, d\gamma_c^s \geq 0. \quad (30)$$

By choosing carefully the discrete Lagrange multiplier component $z_j^\eta$ and test function $\delta z_j^\eta$, one obtains a decoupling of the contact constraints, which yield the set of point-wise conditions

$$\tilde{g}_j \geq 0, \quad z_j^\eta \geq 0, \quad \tilde{g}_j z_j^\eta = 0, \quad (31)$$

in which the weighted gap $\tilde{g}_j$ is defined as

$$\tilde{g}_j \equiv \int_{\{\gamma_c^s\}^h} \Phi_j(\boldsymbol{\xi}^s) \, g^h(\boldsymbol{\xi}^s) \, d\gamma_c^s. \quad (32)$$

The discrete gap function $g^h$ is evaluated as

$$g^h(\boldsymbol{\xi}^s) \equiv \boldsymbol{\eta}^h(\boldsymbol{\xi}^s) \cdot \boldsymbol{g}^h(\boldsymbol{\xi}^s), \quad (33)$$

considering the discrete gap vector in Eq. (29). The discrete frictionless condition simply states that

$$\mathbf{z}_j^\tau = \mathbf{0}. \quad (34)$$

It is worth mentioning that, strictly speaking, the contact constraints expressed via the point-wise conditions in Eq. (31) are only valid for dual interpolation based on the bi-orthogonality and under the assumption of constant unit normal vectors for the node-wise computation, see [29]. This has to do with the localized character of the dual basis functions with regard to the primal variable of the mixed weak formulation.

## 4 Interpolation of the Lagrange multipliers and weighted gap definition

This work heavily relies on the so-called dual mortar method. It consists of defining dual shape functions, herein denoted by $\Phi_j$, to interpolate the Lagrange multipliers satisfying the bi-orthogonality condition,

$$\int_{\{\gamma_c^s\}^h} \Phi_j N_k^s \, d\gamma_c^s = \delta_{jk} \int_{\{\gamma_c^s\}^h} N_k^s \, d\gamma_c^s. \quad (35)$$

This technique becomes particularly advantageous within contact problems, as it localizes the coupling conditions while preserving the optimal convergence of the discretisation error. The first mortar coupling matrix becomes diagonal,

$$\mathbf{D}_{[j,k]} = \mathbf{I}_d \left[ \delta_{jk} \int_{\{\gamma_c^s\}^h} N_j(\boldsymbol{\xi}^s) \, d\gamma_c^s \right], \quad (36)$$

and the contact constraints decouple to point-wise conditions, thus creating a perfect fit for the application of efficient active set strategies. However, when thinking about the requirements of the dual shape functions for contact mechanics, several aspects need to be carefully analysed. More specifically, the inequality nature of contact constraints requires positivity for the Lagrange multiplier basis functions.

First of all, it can be easily shown that dual Lagrange multiplier shape functions are guaranteed to satisfy *partition of unity* on each non-mortar element, i.e.,

$$\sum_{j=1}^{n_s^e} \Phi_j = 1, \quad j = 1, \ldots, n_s^e. \quad (37)$$

This property is assured by the bi-orthogonality condition, see [20] for a proof. Moreover, the partition of unity yields another essential feature of dual Lagrange multiplier basis:

$$\int_{\{\gamma_c^s\}_e^h} \Phi_j \ d\gamma_c^s = \int_{\{\gamma_c^s\}_e^h} N_j^s \ d\gamma_c^s. \quad (38)$$

As mentioned below, this property plays a crucial role in ensuring the integral positivity of Lagrange multiplier interpolation.

While Eqs. (37) and (38) are proven properties of dual shape functions, when considering the transmission of contact stresses across the interface additional requirements are needed. First of all, in order to render the first mortar matrix non-singular, thus invertible, the condition of non-zero integrals arises, i.e.,

$$\int_{\{\gamma_c^s\}_e^h} \Phi_j \ d\gamma_c^s \neq 0. \quad (39)$$

While this requirement is sufficient for mesh tying applications, inequality constraints in contact problems require the Lagrange multiplier shape functions to satisfy at least *integral positivity* further, i.e.,

$$\int_{\{\gamma_c^s\}_e^h} \Phi_j \ d\gamma_c^s > 0. \quad (40)$$

This condition becomes necessary within contact modelling due to the physical interpretation of the HSM conditions. In their discrete form, the HSM conditions are written using the weighted nodal gap $\tilde{g}_j$, which inherits the nodal shape functions used to interpolate the Lagrange multipliers. Therefore, it turns out reasonable to require that positive discrete gap function values, $g^h > 0$, correspond to weighted gaps $\tilde{g}_j$ that are also positive—otherwise, it would lose its physical meaning. From a closer inspection of the definition of the weighted gap, Eq. (32), it is possible to conclude that this is only satisfied if integral positivity of Lagrange multiplier shape functions is guaranteed, i.e., if Eq. (40) holds. Moreover, the first mortar coupling matrix, Eq. (28), is responsible for characterizing the Lagrange multiplier distribution as contact forces on the non-mortar boundary. Negative entries in this matrix can compromise the physical interpretation of the third HSM condition, and, in turn, positivity of $\mathbf{D}_{[j,k]}$ is a critical assumption in the mathematical proof of optimal spatial convergence rates [31].

According to Eq. (38), integral positivity is directly inherited from the corresponding displacement shape functions $N_j^s$. Unfortunately, while for first-order finite element interpolation, this property is fulfilled, for second-order approximation in three dimensions, this aspect reveals to be particularly troublesome. Integral positivity does not hold for specific elements there, e.g., corner nodes of eight-noded quadrilateral (*quad8*) or six-noded triangular (*tri6*) facets.

For standard Lagrange multipliers, in [32], this issue has been addressed either by choosing $n^\lambda < n^s$ (i.e., only the corner nodes carry discrete Lagrange multipliers) or employing *piecewise* linear polynomials on subsegments. In the dual Lagrange multiplier case, a solution based on special basis transformation procedures is proposed in [26,31]. In this approach, a modified basis of interpolation functions based on a well-designed linear combination of the standard interpolation functions $N_j^s$ replaces the latter in the bi-orthogonality condition to construct the corresponding dual shape functions.

### 4.1 The weighted gap and strictly positive shape functions

When dealing with contact problems under large deformations, integral positivity can be viewed as a *minimum* requirement for contact modelling. It assures convergence of the global active set algorithm at least under approximately constant gap conditions, but certainly not in every situation and most likely not for severe gradients found in coarse meshes with high curvatures. An illustrative two-dimensional example with first-order interpolation is represented in Fig. 2, in which the sign of contributions to the weighted gap $\tilde{g}_j$ at a given non-mortar node $j$ is highlighted. Even though there is no overlap between the two boundaries, node $j$ will be erroneously identified as active. The reason for that has to do with the negative part of the dual shape function $\Phi_j$, which yields for those regions weighted gap values $\tilde{g}_j$ with the opposite sign of the discrete gap $g^h$. Whenever severe geometrical curvatures are found, this effect can be amplified and, possibly, yield non-physical results that can compromise the convergence of the active set algorithm.

Since the severity of these artefacts is *h*-dependent, a possible solution to avoid these situations is the refinement of the finite element mesh. However, these options are not always available in practice, due to possible limitations in computational resources or even time to iterate on the numerical model. The ideal solution is a further restriction to *strictly positive* shape functions for the weighted gap interpola-tion, since the possibility of obtaining non-physical contact states is entirely eliminated. However, in the case of strictly positive standard interpolation functions, as in two- and three-dimensional first-order elements, it is mathematically impossible to generate strictly positive dual shape functions. The concept proposed in this work departs from the concept introduced in [25], according to which this problem is solved by changing the shape function used to interpolate the variation of the Lagrange multipliers (i.e. the weighted gap). This means that the dual shape function in Eq. (32) is replaced with the standard shape function, i.e.,

$$\tilde{g}_j \equiv \int_{\{\gamma_c^s\}^h} N_j^s \left(\boldsymbol{\xi}^s\right) g^h \left(\boldsymbol{\xi}^s\right) \, \mathrm{d}\gamma_c^s. \qquad (41)$$

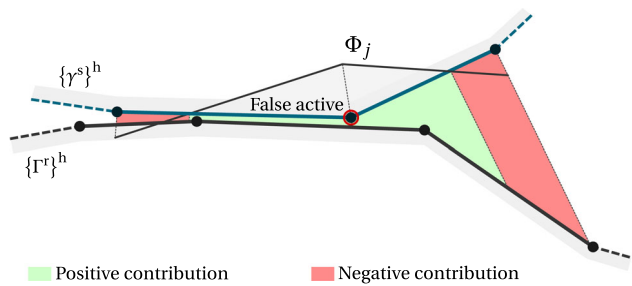

**Fig. 2** Illustration of a possible unphysical contact state in two dimensions

Following this technique, the Lagrange multipliers can still be condensed from the global system of equations, while preserving the robustness of standard Lagrange multipliers for the treatment of the contact constraints.

### 4.2 Additional advantages for rigid/deformable interaction

For the case of rigid/deformable contact, in particular, the application of different functions for the Lagrange multiplier interpolation and weighted gap becomes extremely important for an additional reason: it allows us to completely eliminate the evaluation of the dual shape functions. As already mentioned, the second mortar coupling matrix vanishes due to the fact that one of the opposing contact surfaces remains fixed. Moreover, the bi-orthogonality condition leads to a diagonal first mortar coupling matrix, which can be evaluated using the standard shape functions only, see Eq. (36). Therefore, the only terms left involving the dual shape functions are the contact constraints. As a consequence, and as explained in more detail in the following paragraphs, if the gap function is defined using standard shape functions, the dual shape functions are not explicitly used.

### 4.3 Extension to quadratic elements using piecewise linear interpolation

The establishment of quadratic dual shape functions in three dimensions has always been a challenging topic, especially for contact problems. Additional techniques are required for assuring integral positivity, which, as mentioned previously, is twofold: ensure optimal spatial convergence and physically meaningful weighted gaps. In this work, the locally quadratic technique proposed in [26] is employed to address the first point. Essentially, it consists of combining the bi-orthogonality condition, Eq. (35), with a basis transformation procedure. As suggested in [20], feasible dual shape functions are constructed from the element-wise bi-orthogonality condition, although based on the introduction of modified shape functions $\tilde{N}_j^s$, i.e.,

$$\int_{\{\gamma_c^s\}_e^h} \Phi_j \tilde{N}_k^s \, \mathrm{d}\gamma_c^s = \delta_{jk} \int_{\{\gamma_c^s\}_e^h} \tilde{N}_k^s \, \mathrm{d}\gamma_c^s. \qquad (42)$$

It is noteworthy to mention that this technique leads to a non-diagonal first mortar coupling matrix that, notwithstanding, can still be trivially inverted due to the closed-form character of the transformation scheme. For additional details on this strategy, the reader is referred to the original publication in [26].

If strict positivity of the shape functions used to define the weighted gap is to be further pursued, the situation becomes even more complicated. The original motivation for the technique proposed in [25] relies on the preservation of the properties of standard shape functions in the contact constraints, while keeping the localization character of the dual basis in the coupling conditions. However, it is rather obvious that the requirement of non-negativity is only fully met by standard Lagrange multiplier basis functions for first-order finite element interpolation. For example, Fig. 3 represents the integral value of the quadratic shape function associated with the first node of an 8-noded (serendipity) quadrilateral. The integration domain is considered rectangular, and, as can be observed in the contour plot, there is a region over which the integral value becomes negative. Within the context of the contact formulation, this compromises the physical interpretation of the weighed gap function and impairs the convergence of the active set search. Therefore, an additional modification of the interpolation scheme for the variation of the Lagrange multipliers is required for quadratic elements.

This topic has been firstly addressed in [28] within the context of dual mortar contact with regularisations. Similar to the concepts already introduced in [32], an alternative is proposed to define Lagrange multipliers only at corner nodes, which are interpolated using the associated first-order standard shape functions. Despite leading to a semi-smooth Newton method with a smaller active set to be iterated, this technique has the disadvantage of losing surface information for curved boundaries. The enforcement of contact constraints at edge nodes is ignored and, for coarse meshes, this can lead to non-physical results. Furthermore, as this approach inherently relies on $n^\lambda < n^s$, the construction of the dual basis is slightly more involved and requires substantial algorithmic adaptations, see [26].

In this work, an alternative technique for quadratic finite elements is proposed. In the spirit of the concepts presented in [32] for the evaluation of the mortar integrals for quadratic elements, it is based on the establishment of linear sub-elements and corresponding piecewise interpolation. The weighted gap function is defined using piecewise linear stan-dard shape functions $N_j^{\text{sub}}\left(\boldsymbol{\xi}^{\text{s}}\right)$, i.e.,

$$\tilde{g}_j \equiv \int_{\{\gamma_c^s\}^h} N_j^{\text{sub}}\left(\boldsymbol{\xi}^{\text{s}}\right) g^{\text{h}}\left(\boldsymbol{\xi}^{\text{s}}\right) \, d\gamma_c^s. \tag{43}$$

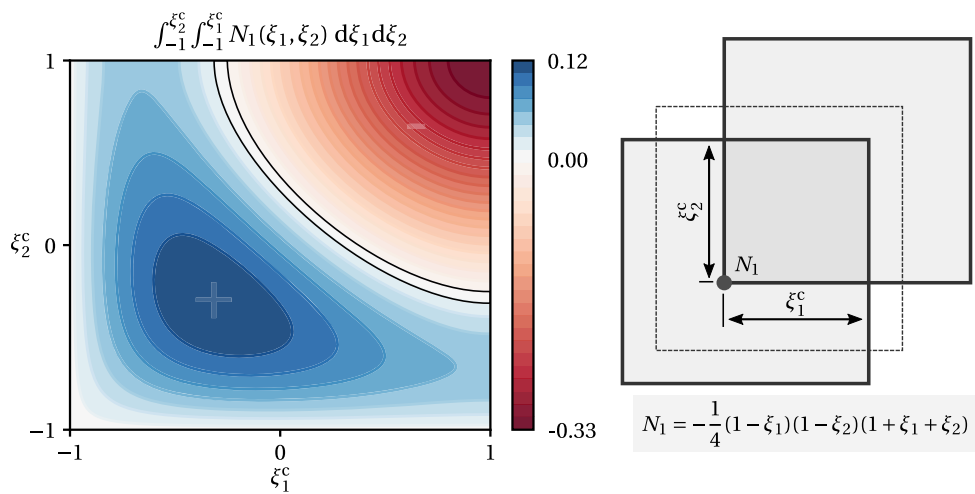

**Fig. 3** The integral value of the standard shape function node 1, $N_1$, of the 8-noded quadratic element over a partial rectangular domains

Compared with the alternative proposed by [28], all non-mortar nodes are checked for contact, thus preserving more information about the interface geometry. As the condition $n^\lambda = n^s$ is retained, the implementation of the proposed approach requires little effort and enables recycling of most of the pre-existing algorithmic components. In combination with the certainty of strict positivity of the piecewise linear shape functions, this allows for a robust algorithm that becomes less sensitive to discretization problems. However, as a trade-off, the computational complexity in the numerical evaluation of mortar integrals increases. Instead of approximating a single linear element (containing only the corner nodes), each quadratic facet is divided into multiple linear/bilinear sub-elements, which may require further sub-divisions within the clipping polygon algorithm. This downside, however, is not intrinsically related to the proposed approach but is already present in popular mortar segmentation procedures for quadratic elements in deformable/deformable contact [32]. Notwithstanding, for rigid/deformable contact, this aspect can be counterbalanced by employing the efficient projected orthonormal frame to be described in Sect. 5. The application of the piecewise linear interpolation requires the establishment of proper mappings, and we refer to Appendix A for their explicit expressions and associated Jacobian matrices. Table 1 gives an overview of the various finite element interpolation schemes used within the proposed algorithm, also representing the shape functions characteristic of both corner and edge nodes of a quadratic 8-noded quadrilateral (quad8).

## 5 Projected averaged moving frame

With the discrete version of the contact virtual work and contact constraints thought out, attention is now shifted towards their computational treatment. The importance of mortar integral evaluation within mortar methods is well-known and represents one of the main challenges. Their correct evaluation is essential to preserve the sought-after properties of the mortar-based variational formulation. It requires approximating surface integrals with complex geometrical operations involved and, therefore, ends up being one of the main contributors to the overall cost of the algorithm. Therefore, there is a clear motivation for new strategies that are able to facilitate the complexity of the algorithm without affecting its accuracy and robustness. For the particular case at hand of rigid/deformable contact under large deformations, the fact that one of the boundaries remains fixed can be exploited and, as explained in the following paragraphs, a new definition for the continuous field of orthonormal frames is proposed.

In all contact problems, it is necessary to define a local orthonormal moving frame attached to each contact node containing a Lagrange multiplier. It splits the surface contact tractions into normal and tangential components and establishes projection rules necessary for the evaluation of mortar coupling terms. The overall idea of the proposed method consists in defining an initial field of orthonormal frames on the rigid side, which is then continuously projected onto the non-mortar side throughout the deformation process. On the ground foundation of this idea is the fact that, while contact occurs, the boundaries of both sides tend to coincide and, for the regions in full contact, become practically identical. Therefore, the field of orthonormal frames on the rigid side can be defined using sophisticated and accurate methodologies, which are then transmitted to the deformable side by means of simple projections rules, similar to the ones already

**Table 1** Finite element interpolations for rigid/deformable dual mortar contact and shape functions for a quadratic 8-noded quadrilateral

**(a) Standard shape functions**

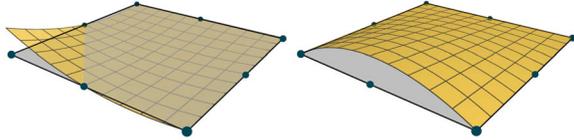

Geometry
$$x^r \approx \{x^r\}^h \equiv \sum_{k=1}^{n^s} N_k^s(\xi^s)\, x_k^s$$
$$x^s \approx \{x^s\}^h \equiv \sum_{k=1}^{n^s} N_k^s(\xi^s)\, x_k^s$$

Displacements
$$u^s \approx \{u^s\}^h \equiv \sum_{k=1}^{n^s} N_k^s(\xi^s)\, d_k^s$$

Displacements variation
$$\delta u^s \approx \{\delta u^s\}^h \equiv \sum_{k=1}^{n^s} N_k^s(\xi^s)\, \delta d_k^s$$

**(b) Dual shape functions**

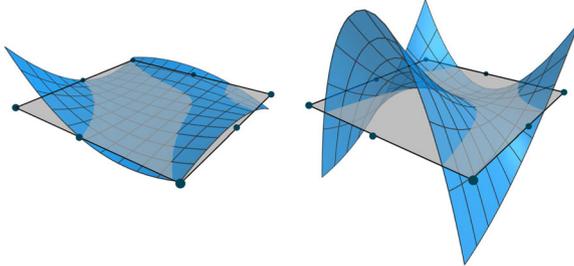

Lagrange multipliers
$$\lambda \approx \lambda^h \equiv \sum_{j=1}^{n^s} \Phi_j(\xi^s)\, z_j$$

**(c) Piecewise linear shape functions**

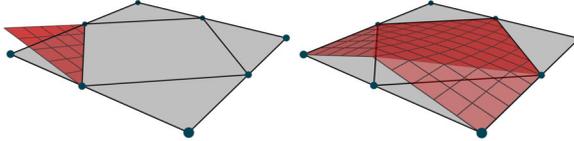

Weighted gap function
$$\tilde{g}_j \equiv \int_{\{\gamma_c^s\}^h} N_j^{\text{sub}}(\xi^s)\, g^h(\xi^s)\, d\gamma_c^s$$

used within the integration algorithm. The individual steps are described in more detail in the following.

Firstly, it is necessary to define the field of orthonormal frames on the rigid side. This operation is only realised once during problem initialisation and the averaged normal approach is employed here. Originally proposed by [12], and later on slightly simplified by [22], it is based on a continuous field of normal vectors defined on the non-mortar side, which smooths the discontinuities associated with the discretization of the contact interface by averaging the nodal unit normals.

Next, the unit normal vector is projected to the non-mortar contact boundary and inverted in order to still point outwards, see Fig. 4. For a given non-mortar node $j$ with coordinates $\mathbf{x}_j^s$, the projection consists of finding the isoparametric coordinate $\hat{\boldsymbol{\xi}}^r$ on the rigid side such that the following condition holds:

$$\sum_{k=1}^{n_e^r} N_k^r(\hat{\boldsymbol{\xi}}^r)\left[\mathbf{x}_k^r + \alpha\tilde{\boldsymbol{\eta}}_k^r\right] - \mathbf{x}_j^s = \mathbf{0}. \tag{44}$$

Here, $n_e^r$ denotes the total number of nodes of the rigid element and $\tilde{\boldsymbol{\eta}}_k^r$ the averaged unit normal vectors. The parameter $\alpha$ relates to the normal distance between the points.[1] This system of equations can be solved with a local Newton–Raphson procedure, in which each iteration reads

$$\begin{Bmatrix} \Delta\hat{\xi}_1^r \\ \Delta\hat{\xi}_2^r \\ \Delta\alpha \end{Bmatrix} = [\mathbf{W}]^{-1} \left\{ \sum_{l=1}^{n_e^r} N_l^r(\hat{\boldsymbol{\xi}}^r)\left(\mathbf{x}_l^r + \alpha\tilde{\boldsymbol{\eta}}_k^r\right) - \mathbf{x}_j^s \right\}. \tag{45}$$

The matrix $[\mathbf{W}] \in \mathbb{R}^{d\times d}$ is obtained from the derivative of the projection condition with respect to the rigid side coordinate $\boldsymbol{\xi}^r$ and the parameter $\alpha$. In three dimensions, it yields

$$[\mathbf{W}] \equiv \left[ \sum_{l=1}^{n_e^r} N_{l,\xi_1^r}^r\left(\mathbf{x}_l^r + \tilde{\boldsymbol{\eta}}_l\right) \;\Big|\; \sum_{l=1}^{n_e^r} N_{l,\xi_2^r}^r\left(\mathbf{x}_l^r + \tilde{\boldsymbol{\eta}}_l\right) \;\Big|\; \sum_{l=1}^{n_e^r} N_l^r\, \tilde{\boldsymbol{\eta}}_l \right]. \tag{46}$$

The two-dimensional version is relatively straightforward by simply omitting the second column. Lastly, with the projection coordinate $\hat{\boldsymbol{\xi}}^r$ at hand, the associated frame can be

---

[1] For the sake of simplicity, the interpolated normal vector is not normalized. This does not change the solution of the projection procedure, thus affecting the physical meaning of the parameter $\alpha$ only.

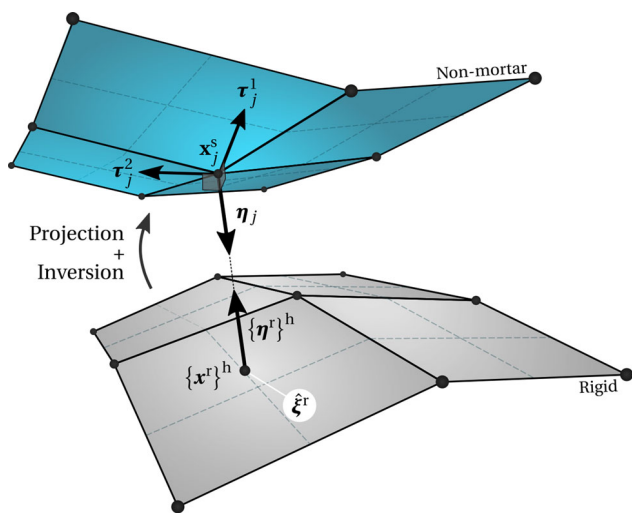

**Fig. 4** Projection of the averaged orthonormal frame from the rigid onto the non-mortar boundary

inverted, which means that the unit normal vector $\eta_j^h$ at the non-mortar node $j$ yields

$$\eta_j^h = -\frac{\sum_{l=1}^{n_e^r} N_l^r(\hat{\xi}^r)\tilde{\eta}_k^r}{\left\|\sum_{l=1}^{n_e^r} N_l^r(\hat{\xi}^r)\tilde{\eta}_k^r\right\|}. \tag{47}$$

Having the unit normal vector at hand, the tangent vector can be freely chosen from the tangential plane. A recommended technique is, for example, considering the direction of the interpolated tangent vector on the rigid side. In practice, the projection is performed considering various rigid interface elements until a valid solution is found. The application of an efficient global search algorithm is, thus, highly recommended in order to perform this iterative procedure based on a reduced list of target elements.

*Remark 1* In situations where no valid projection is found (e.g., in dropping edge problems), the normal vector on the non-mortar boundary can be defined using the averaging technique locally, as it depends only on the deformable boundary itself.

The local frame is deformation-dependent for contact under large deformations, and, therefore, needs to be linearised within the Newton–Raphson algorithm. In fact, this now exposes the main advantage of this technique, which regards the self-contained character of the projection procedure. As demonstrated in more detail in Sect. 7.1, the projected frame derivatives are guaranteed to have the minimum bandwidth, i.e., they contain only the degrees of freedom associated with the non-mortar node itself. This has a beneficial impact on the computational complexity of the algorithm by reducing the total number of individual operations. As the derivatives of the normal and tangent vectors appear (directly and indirectly) in every term of the formulation, the reduction in the number of terms in the derivatives becomes amplified in the overall computational cost. Recall that, beyond the contact constraints, the integration scheme is based on projections that also use the continuous field of normals on the non-mortar side. The improvements in computational complexity are carefully investigated and quantitatively measured in the numerical examples shown in Sect. 8.2.

## 6 Numerical evaluation of mortar integrals

The prime cause for the difficulty in evaluating mortar integrals is related to the quantities with terms belonging to different sides of the contact interface—usually connected by means of projection rules that, within the ongoing formulation, are based on the projected normals. Generally speaking, these include the transmission of contact stresses across the interface, described by the second mortar coupling matrix, and the kinematics describing the relative motion between both boundaries, namely in the form of the gap vector. One of the main simplifications for the particular case of rigid/deformable contact is that the second mortar matrix vanishes. However, even in the simplest case of frictionless contact, the weighted gap still needs to be evaluated and, therefore, the challenge of correctly evaluating mortar integrals is still present.

The primary source of complexity involves the transmission of geometrical information between boundaries (to determine the overlap of both domains) and, in the three-dimensional case, the evaluation of surface integrals with complex geometries. As the rigid boundary is interpolated using finite elements, the techniques described in the overview work by [30] are equally applicable to the problem at hand. Both the segmentation and element-wise integration schemes remain practically unchanged, with the only exception being the displacements of the mortar side vanishing. Nonetheless, for quadratic interpolation using piecewise linear interpolation, the understanding of how the subdivision of the interface element into multiple sub-elements affects the integration scheme is noteworthy to mention. In the spirit of preserving all the techniques derived for linear elements, each individual sub-element of the parent quadratic facet is treated accordingly. The two strategies employed for the numerical approximation of mortar integrals are described in the following.

### 6.1 Evaluation of mortar integrals for piecewise linear interpolation

For the segmentation strategy in two dimensions, segments are formed using each pair of nodes of the involved 3-noded line elements, while in three dimensions, the clipping polygon is established using the sub-elements (3-noded triangles or 4-noded quadrilaterals). Because the clipping polygon technique for three-dimensional problems is only valid for linear facets (otherwise, it would be impractical to perform the clipping algorithm based on curved domains), this means that the quadratic facets on the rigid side also need to be divided into sub-elements. The basic steps of the algorithm remain the same: after defining the clipping polygon, it is divided into multiple cells to be numerically integrated (triangles and quadrilaterals) and the Gauss points are projected back to the sub-element. At this stage, however, an additional step needs to be performed by recovering the original isoparametric coordinate at the parent element (quadratic) using the mappings described in Appendix A. The contribution $D_{jj}$ of a given pair of non-mortar and rigid elements to the first mortar coupling matrix, see Eq. (36), becomes

$$D_{jj} \approx \sum_{s=1}^{n_{\text{sub}}} \sum_{c=1}^{n_{\text{cell}}} \sum_{g=1}^{n_g} w_g \, N_j\bigl(\boldsymbol{\xi}_g^{\text{s}}\bigl(\boldsymbol{\xi}_g^{\text{sub}}(\boldsymbol{\zeta}_g)\bigr)\bigr) J_c, \quad (48)$$

where $n_{\text{sub}}$ stands for the total number of sub-elements and $n_{\text{cell}}$ is the total number of integration cells. The Gaussian quadrature is defined over $n_g$ integration points with $w_g$ weights and coordinates $\boldsymbol{\zeta}_g$. The Jacobian determinant, $J_c$, defines the transformation from the integration cell of the sub-element to the global spatial configuration, i.e.,

$$J_c\bigl(\boldsymbol{\xi}^{\text{sub}}\bigr) = \left\| \frac{\partial \{x^s\}^h}{\partial \boldsymbol{\xi}^s} \right\| \cdot \left\| \frac{\partial \boldsymbol{\xi}^s}{\partial \boldsymbol{\xi}^{\text{sub}}} \right\|. \quad (49)$$

For the element-based integration, the Gauss points are defined on the sub-element and directly projected to the opposite side. Figure 5 schematically represents the boundary-segmentation method in three dimensions, which is based on the combination of both strategies. The algorithm is illustrated for a pair of non-mortar and rigid elements, which, after division into sub-elements, yields both types of integration cells. For the sub-element represented in red, all the Gauss points are successfully projected, whereas the remaining sub-elements require segmentation.

It is important to mention that, from a practical perspective, there exists a slight difference in both strategies that motivates a modified physical interpretation of the Jacobian determinant. For element-wise integration, all the nodes of the integration cell belong to the parent finite element and, therefore, the mapping to the parent element is explicitly employed and the Jacobian evaluated using the two terms in Eq. (49). However, when applying the segmentation scheme, the sub-element is projected to the auxiliary plane, which leads to a relative loss of geometrical information (inter-preted as a kind of faceting of the quadratic element).[2] The nodes of the integration cells become either projections of the sub-element nodes to the auxiliary plane or nodes generated by the clipping algorithm. The Jacobian determinant is evaluated directly by using the global coordinates of the integration cell nodes and, thus, overlooking the two-step operation in Eq. (49).

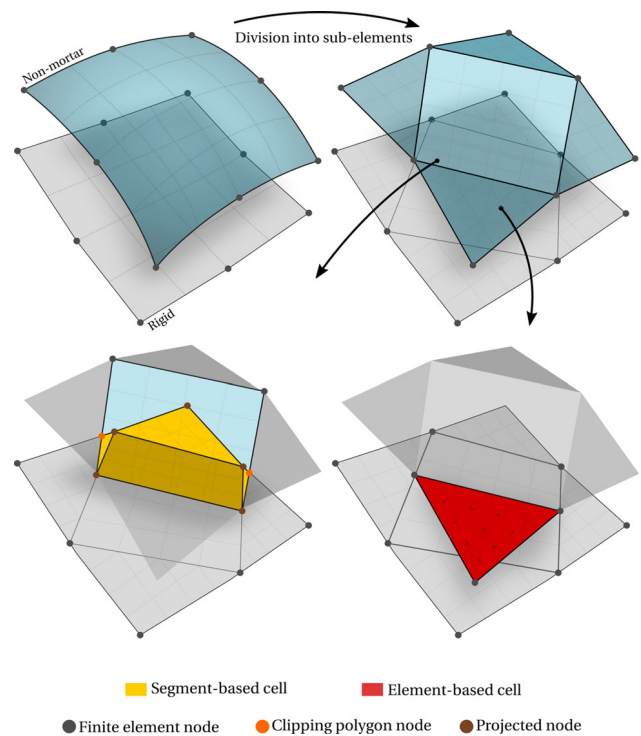

**Fig. 5** Schematic illustration of the boundary-segmentation integration method for the piecewise linear interpolation in three dimensions. The top two figures represent the division into sub-elements and the bottom two figures correspond to the integration cells originated from two of the five sub-elements

## 7 Global solution algorithm

The final step to obtain the final discrete non-linear contact problem between a deformable body and a rigid obstacle regards the active set strategy. As for the general case of unilateral contact between two deformable bodies, the contact inequality constraints require identifying the subset of interface nodes effectively in contact. To address this problem, the primal-dual active set strategy (PDASS) described in [18] is thoroughly applied without any modifications. In a nutshell, it consists of reformulating the discrete nodal inequality constraints using a non-linear complementarity (NCP) function. This introduces a certain regularisation on the active set search and enables the application of a Newton–Raphson type algorithm, comprising not only geometrical and material non-linearities, but also the contact active set search itself. The final discrete contact problem can be written as the entire set of equality conditions:

$$\mathbf{r}(\mathbf{d}, \mathbf{z}) \equiv \mathbf{f}_{\text{int}}(\mathbf{d}) - \mathbf{f}_{\text{ext}} + \mathbf{f}_{\text{c}}(\mathbf{d}, \mathbf{z}) = \mathbf{0}, \qquad (50)$$

$$z_j^{\eta} - \max\left\{0, z_j^{\eta} - c^{\eta} \tilde{g}_j\right\} = 0, \quad j = 1, \dots, n^{\text{s}}, \qquad (51)$$

$$\mathbf{z}_j^{\tau} = \mathbf{0}, \quad j = 1, \dots, n^{\text{s}}. \qquad (52)$$

---

[2] Strictly speaking, this becomes relevant only for 4-noded quadrilateral facets, in which the possibility of element warping exists.

Recall that $\mathbf{r}(\mathbf{d}, \mathbf{z})$ represents the residual vector, $\mathbf{f}_{\text{int}}(\mathbf{d})$ and $\mathbf{f}_{\text{ext}}$ the internal and external forces vectors (which remain unchanged from classical finite element problems). The vector $\mathbf{f}_{\text{c}}(\mathbf{d}, \mathbf{z})$ stands for the discrete contact forces. At this stage, the foundations for the application of a Newton-type algorithm are complete. Thus, we describe in the following the consistent linearisation of the problem.

## 7.1 Consistent linearisation

The application of the semi-smooth Newton algorithm requires consistent linearisation of both the discrete balance equations and the NCP functions. In this section, the focus is exclusively placed on the terms introduced within the current formulation for non-linear rigid/deformable contact, namely the piecewise linear interpolation for quadratic dual mortar contact and the projected orthonormal frame. The remaining derivations remain unchanged from the unilateral contact case and, therefore, the reader is referred to the discussions in [11,22,33], to name a few. In what follows, the compact notation $\mathscr{D}(\bullet)$ stands for the so-called directional derivative at a given iteration k, i.e.

$$\mathscr{D}(\bullet) \equiv \left.\frac{\partial(\bullet)}{\partial \mathbf{d}}\right|^k \Delta \mathbf{d} + \left.\frac{\partial(\bullet)}{\partial \mathbf{z}}\right|^k \Delta \mathbf{z}. \qquad (53)$$

The vectors $\Delta \mathbf{d}$ and $\Delta \mathbf{z}$ stand for the solution increment in the displacements and Lagrange multipliers, respectively.

### 7.1.1 Piecewise linear interpolation

The piecewise linear interpolation affects the definition of the isoparametric coordinates at each Gauss point. As the interface element is divided into sub-elements, the numerical integration requires the application of mappings between domains. These need to be taken into account within the derivative chain rule and, for example, the derivative of the isoparametric non-mortar coordinate, $\mathscr{D}\boldsymbol{\xi}^{\text{s}}$, yields

$$\mathscr{D}\boldsymbol{\xi}^{\text{s}} = \frac{\partial \boldsymbol{\xi}^{\text{s}}}{\partial \boldsymbol{\xi}^{\text{sub}}} \mathscr{D}\boldsymbol{\xi}^{\text{sub}}. \qquad (54)$$

The first term regards the Jacobian matrix of the mappings, and we refer to Appendix A for its definition. The derivative $\mathscr{D}\boldsymbol{\xi}^{\text{sub}}$ contains the directional derivative of the sub-element parameter space, which is computed using the typical first-order element procedures.

### 7.1.2 Projected orthonormal frame

Considering the definition of the projected unit normal given in Eq. (47), its linearisation reads

$$\mathscr{D}\boldsymbol{\eta}_j^{\text{h}} = \left(\frac{1}{l_{\check{\boldsymbol{\eta}}^{\text{r}}}}\boldsymbol{I} - \frac{1}{l_{\check{\boldsymbol{\eta}}^{\text{r}}}^3}\check{\boldsymbol{\eta}}^{\text{r}} \otimes \check{\boldsymbol{\eta}}^{\text{r}}\right)\mathscr{D}\check{\boldsymbol{\eta}}^{\text{r}}. \qquad (55)$$

Here, $\check{\boldsymbol{\eta}}_j^{\text{r}}(\hat{\boldsymbol{\xi}}^{\text{r}})$ has been introduced as an abbreviation for the non-unit normal vector on the rigid boundary,

$$\check{\boldsymbol{\eta}}_j^{\text{r}}(\hat{\boldsymbol{\xi}}^{\text{r}}) \equiv \sum_{l=1}^{n_{\text{e}}^{\text{r}}} N_l^{\text{r}}(\hat{\boldsymbol{\xi}}^{\text{r}}) \tilde{\boldsymbol{\eta}}_k^{\text{r}}, \qquad (56)$$

of length $l_{\check{\boldsymbol{\eta}}^{\text{r}}}$ and whose derivative yields

$$\mathscr{D}\check{\boldsymbol{\eta}}_j^{\text{r}}(\hat{\boldsymbol{\xi}}^{\text{r}}) = \sum_{k=1}^{n_{\text{e}}^{\text{r}}} N_{k,\xi_1^{\text{r}}}^{\text{r}}(\hat{\boldsymbol{\xi}}^{\text{r}}) \mathscr{D}\hat{\xi}_1^{\text{r}}\, \tilde{\boldsymbol{\eta}}_k^{\text{r}}$$
$$+ \sum_{k=1}^{n_{\text{e}}^{\text{r}}} N_{k,\xi_2^{\text{r}}}^{\text{r}}(\hat{\boldsymbol{\xi}}^{\text{r}}) \mathscr{D}\hat{\xi}_2^{\text{r}}\, \tilde{\boldsymbol{\eta}}_k^{\text{r}}. \qquad (57)$$

Note that the only derivative needed arises from the projection procedure, given that the averaged unit normal linearisation vanishes (the rigid boundary remains fixed). The linearisation of the projected rigid isoparametric coordinate $\hat{\boldsymbol{\xi}}^{\text{r}}$ can be extracted from the projection condition in Eq. (45) as follows:

$$\begin{Bmatrix} \mathscr{D}\hat{\xi}_1^{\text{r}} \\ \mathscr{D}\hat{\xi}_2^{\text{r}} \\ \mathscr{D}\alpha \end{Bmatrix} = [\mathbf{W}]^{-1}\{\Delta \mathbf{x}_j^{\text{s}}\}. \qquad (58)$$

It is noteworthy to mention that the matrix $[\mathbf{W}]$ to be inverted is already computed during the projection, see Eq. (46), which makes the computational evaluation of the orthonormal moving frame relatively straightforward and efficient.

## 7.2 Algebraic representation

This last section provides the algebraic representation of the discrete entities involved in the contact algorithm. All nodes and corresponding degrees of freedom are partitioned into two (instead of three) disjoint sets $\mathcal{S} \cup \mathcal{N}$: a group $\mathcal{S}$ containing all non-mortar quantities and a group $\mathcal{N}$ associated with all remaining nodes or degrees of freedom. Then, the non-mortar set $\mathcal{S}$ is further partitioned into two disjoint sets: the inactive nodes set $\mathcal{I}$ the set $\mathcal{A}$ of nodes in contact. The assembled system to be solved within each semi-smooth Newton step k in order to obtain the incremental displacements vector $\Delta\mathbf{d}$ and current Lagrange multipliers $\mathbf{z}^{k+1}$ can be expressed as:

$$\begin{bmatrix} \mathbf{K}_{\mathcal{N}\mathcal{N}} & \mathbf{K}_{\mathcal{N}\mathcal{I}} & \mathbf{K}_{\mathcal{N}\mathcal{A}} & \mathbf{0} & \mathbf{0} \\ \mathbf{K}_{\mathcal{I}\mathcal{N}} & \widetilde{\mathbf{K}}_{\mathcal{I}\mathcal{I}} & \widetilde{\mathbf{K}}_{\mathcal{I}\mathcal{A}} & \mathbf{D}^{\mathrm{T}}_{\mathcal{I}\mathcal{I}} & \mathbf{D}^{\mathrm{T}}_{\mathcal{I}\mathcal{A}} \\ \mathbf{K}_{\mathcal{A}\mathcal{N}} & \widetilde{\mathbf{K}}_{\mathcal{A}\mathcal{I}} & \widetilde{\mathbf{K}}_{\mathcal{A}\mathcal{A}} & \mathbf{D}^{\mathrm{T}}_{\mathcal{A}\mathcal{I}} & \mathbf{D}^{\mathrm{T}}_{\mathcal{A}\mathcal{A}} \\ \mathbf{0} & \mathbf{0} & \mathbf{0} & \mathbf{I}_{\mathcal{I}} & \mathbf{0} \\ \mathbf{0} & \mathbf{A}_{\mathcal{I}} & \mathbf{A}_{\mathcal{A}} & \mathbf{0} & \mathbf{0} \\ \mathbf{0} & \mathbf{F}_{\mathcal{I}} & \mathbf{F}_{\mathcal{A}} & \mathbf{0} & \mathbf{T} \end{bmatrix} \begin{Bmatrix} \Delta\mathbf{d}_{\mathcal{N}} \\ \Delta\mathbf{d}_{\mathcal{I}} \\ \Delta\mathbf{d}_{\mathcal{A}} \\ \mathbf{z}^{k+1}_{\mathcal{I}} \\ \mathbf{z}^{k+1}_{\mathcal{A}} \end{Bmatrix} = - \begin{Bmatrix} \mathbf{r}_{\mathcal{N}} \\ \tilde{\mathbf{r}}_{\mathcal{I}} \\ \tilde{\mathbf{r}}_{\mathcal{A}} \\ \mathbf{0} \\ \tilde{\mathbf{g}} \\ \mathbf{0} \end{Bmatrix}. \quad (59)$$

The blocks $\mathbf{K}$ denote the stiffness matrix resulting from the linearisation of the internal forces vector. The blocks $\widetilde{\mathbf{K}}$ represent the effective stiffness matrix, obtained from the summation of the respective stiffness blocks $\mathbf{K}$ with the linearization terms of the contact force vector $\mathbf{f}_c$ with respect to the displacements, i.e.,

$$[\widetilde{\mathbf{K}}(\mathbf{d},\mathbf{z})] \equiv [\mathbf{K}(\mathbf{d})] + \begin{bmatrix} \mathbf{0} \\ \mathscr{D}\mathbf{D}^{\mathrm{T}} \end{bmatrix} \{\mathbf{z}\}. \quad (60)$$

The matrix blocks of $\mathbf{D}$ are abbreviated such that, for instance,

$$\mathbf{D}^{\mathrm{T}}_{\mathcal{I}\mathcal{A}} \equiv [\mathbf{D}^{\mathrm{T}}]_{\mathcal{I}\mathcal{A}}. \quad (61)$$

Notice that

$$\mathbf{D}^{\mathrm{T}}_{\mathcal{I}\mathcal{A}} \neq [\mathbf{D}_{\mathcal{I}\mathcal{A}}]^{\mathrm{T}}. \quad (62)$$

The blocks $\mathbf{A}$ contain the derivatives with respect to the displacements of the NCP function for the normal contact constraints. The matrices $\mathbf{F}$ and $\mathbf{T}$ aggregates the derivative of the frictionless condition (52) with respect to the displacements and Lagrange multiplier, accordingly. The residual blocks $\tilde{\mathbf{r}}$ stand for the abbreviation $\tilde{\mathbf{r}} \equiv \mathbf{f}_{\mathrm{int}} - \mathbf{f}_{\mathrm{ext}}$, which comes as a result of solving directly for the unknown Lagrange multipliers $\mathbf{z}^{k+1}$ at each iteration (i.e., without employing an incremental formulation). The vector $\tilde{\mathbf{g}}$ gathers all the nodal weighted gaps. Compared with the counterpart for unilateral contact, the only difference is that there are no mortar degrees of freedom, thus leading to a reduced number of stiffness matrix blocks.

**Remark 2** It should be mentioned that, in Eq. (59), the case of a non-diagonal first mortar coupling matrix is assumed for the sake of generality. Nonetheless, for first-order finite elements, it retains the diagonal structure due to the bi-orthogonality condition given in Eq. (35). For second-order interpolation, on the other hand, the application of the modified dual shape functions, see Eq. (42), leads to a non-diagonal first mortar matrix (although, still easily inverted).

### 7.2.1 Elimination of the Lagrange multipliers

As already mentioned, the use of dual Lagrange multipliers allows for a straightforward simplification of the system of equations by performing the condensation of the Lagrange multipliers (thus removing the unwanted saddle point structure). This is possible because the first mortar coupling matrix $\mathbf{D}$ is trivially inverted. The evaluation of the Lagrange multipliers at a given configuration starts with the consideration of the following system of equations

$$\begin{Bmatrix} \mathbf{z}^{k+1}_{\mathcal{I}} \\ \mathbf{z}^{k+1}_{\mathcal{A}} \end{Bmatrix} = - \begin{bmatrix} \mathbf{D}^{-\mathrm{T}}_{\mathcal{I}\mathcal{I}} & \mathbf{D}^{-\mathrm{T}}_{\mathcal{I}\mathcal{A}} \\ \mathbf{D}^{-\mathrm{T}}_{\mathcal{A}\mathcal{I}} & \mathbf{D}^{-\mathrm{T}}_{\mathcal{A}\mathcal{A}} \end{bmatrix} \begin{Bmatrix} \tilde{\mathbf{r}}_{\mathcal{I}} + \sum_{\mathcal{X}\in\{\mathcal{N},\mathcal{I},\mathcal{A}\}} \mathbf{K}_{\mathcal{I}\mathcal{X}} \Delta\mathbf{d}_{\mathcal{X}} \\ \tilde{\mathbf{r}}_{\mathcal{A}} + \sum_{\mathcal{X}\in\{\mathcal{N},\mathcal{I},\mathcal{A}\}} \mathbf{K}_{\mathcal{A}\mathcal{X}} \Delta\mathbf{d}_{\mathcal{X}} \end{Bmatrix}. \quad (63)$$

Consideration of the fourth row of Eq. (59) yields $\mathbf{z}^{k+1}_{\mathcal{I}} = \mathbf{0}$ and, therefore, one has

$$\mathbf{z}^{k+1}_{\mathcal{A}} = - \sum_{\mathcal{Y}\in\{\mathcal{I},\mathcal{A}\}} \mathbf{D}^{-\mathrm{T}}_{\mathcal{A}\mathcal{Y}} \left[ \tilde{\mathbf{r}}_{\mathcal{Y}} + \sum_{\mathcal{X}\in\{\mathcal{N},\mathcal{I},\mathcal{A}\}} \mathbf{K}_{\mathcal{Y}\mathcal{X}} \Delta\mathbf{d}_{\mathcal{X}} \right]. \quad (64)$$

This allows us also to solve the third row of Eq. (59). It is important to highlight that, in the equations above, the blocks of the inverse of the first mortar coupling matrix are obtained from the global inverse, i.e.,

$$\mathbf{D}^{-\mathrm{T}}_{\mathcal{I}\mathcal{A}} \equiv [\mathbf{D}^{-\mathrm{T}}]_{\mathcal{I}\mathcal{A}}. \quad (65)$$

Notice that

$$\mathbf{D}^{-\mathrm{T}}_{\mathcal{I}\mathcal{A}} \neq [\mathbf{D}_{\mathcal{I}\mathcal{A}}]^{-\mathrm{T}}. \quad (66)$$

It should be also mentioned that the presented condensation procedure becomes simplified for first-order finite elements, as the first mortar coupling matrix maintains a diagonal structure.

Substitution of the expressions above into Eq. (59) leads to the final condensed system:

$$\begin{bmatrix} \mathbf{K}_{\mathcal{N}\mathcal{N}} & \mathbf{K}_{\mathcal{N}\mathcal{I}} & \mathbf{K}_{\mathcal{N}\mathcal{A}} \\ \check{\mathbf{K}}_{\mathcal{N}} & \check{\mathbf{K}}_{\mathcal{I}} & \check{\mathbf{K}}_{\mathcal{A}} \\ \mathbf{0} & \mathbf{A}_{\mathcal{I}} & \mathbf{A}_{\mathcal{Q}} \\ \check{\mathbf{F}}_{\mathcal{N}} & \check{\mathbf{F}}_{\mathcal{I}} & \check{\mathbf{F}}_{\mathcal{A}} \end{bmatrix} \begin{Bmatrix} \Delta\mathbf{d}_{\mathcal{N}} \\ \Delta\mathbf{d}_{\mathcal{I}} \\ \Delta\mathbf{d}_{\mathcal{A}} \end{Bmatrix} = - \begin{Bmatrix} \mathbf{r}_{\mathcal{N}} \\ \check{\mathbf{r}}_{\mathcal{I}} \\ \tilde{\mathbf{g}} \\ \check{\mathbf{r}}_{\mathcal{A}} \end{Bmatrix}. \quad (67)$$

Here, several algebraic functions and abbreviations have been introduced to facilitate the notation. The algebraic function $\check{\mathbf{K}}_{(\bullet)}$ is defined as

$$\check{\mathbf{K}}_{(\bullet)} \equiv \begin{cases} \mathbf{K}_{\mathcal{I}(\bullet)} - \sum_{\mathcal{Y} \in \{\mathcal{I},\mathcal{A}\}} \mathbf{D}_{\mathcal{I}\mathcal{A}}^{\mathrm{T}} \mathbf{D}_{\mathcal{A}\mathcal{Y}}^{-\mathrm{T}} \mathbf{K}_{\mathcal{Y}\mathcal{N}} \Delta \mathbf{d}_{\mathcal{N}}, & \text{if } (\bullet) \in \mathcal{N}, \\ \widetilde{\mathbf{K}}_{\mathcal{I}(\bullet)} - \sum_{\mathcal{Y} \in \{\mathcal{I},\mathcal{A}\}} \mathbf{D}_{\mathcal{I}\mathcal{A}}^{\mathrm{T}} \mathbf{D}_{\mathcal{A}\mathcal{Y}}^{-\mathrm{T}} \mathbf{K}_{\mathcal{Y}(\bullet)} \Delta \mathbf{d}_{(\bullet)}, & \text{if } (\bullet) \notin \mathcal{N}, \end{cases}$$
(68)

and the corresponding residual comes as

$$\check{\mathbf{r}}_{\mathcal{I}} \equiv \tilde{\mathbf{r}}_{\mathcal{I}} - \sum_{\mathcal{Y} \in \{\mathcal{I},\mathcal{A}\}} \mathbf{D}_{\mathcal{I}\mathcal{A}}^{\mathrm{T}} \mathbf{D}_{\mathcal{A}\mathcal{Y}}^{-\mathrm{T}} \tilde{\mathbf{r}}_{\mathcal{Y}}.$$
(69)

The algebraic function $\check{\mathbf{F}}_{(\bullet)}$ reads

$$\check{\mathbf{F}}_{(\bullet)} \equiv \begin{cases} -\sum_{\mathcal{Y} \in \{\mathcal{I},\mathcal{A}\}} \mathbf{T} \mathbf{D}_{\mathcal{A}\mathcal{Y}}^{-\mathrm{T}} \mathbf{K}_{\mathcal{Y}\mathcal{N}} \Delta \mathbf{d}_{\mathcal{N}}, & \text{if } (\bullet) \in \mathcal{N}, \\ \mathbf{F}_{(\bullet)} - \sum_{\mathcal{Y} \in \{\mathcal{I},\mathcal{A}\}} \mathbf{T} \mathbf{D}_{\mathcal{A}\mathcal{Y}}^{-\mathrm{T}} \mathbf{K}_{\mathcal{Y}(\bullet)} \Delta \mathbf{d}_{(\bullet)}, & \text{if } (\bullet) \notin \mathcal{N}, \end{cases}$$
(70)

and the corresponding residual $\check{\mathbf{r}}_{\mathcal{A}}$ yields

$$\check{\mathbf{r}}_{\mathcal{A}} \equiv -\sum_{\mathcal{Y} \in \{\mathcal{I},\mathcal{A}\}} \mathbf{T} \mathbf{D}_{\mathcal{A}\mathcal{Y}}^{-\mathrm{T}} \tilde{\mathbf{r}}_{\mathcal{Y}}.$$
(71)

# 8 Numerical results

In the following, several numerical examples are presented and analysed in order to validate the proposed formulation for rigid/deformable finite deformation contact. The set of numerical examples presented is exclusively focused on particular aspects of the proposed formulation. Firstly, the optimal convergence rate of the formulation using the modified weighted gap function is discussed in Sect. 8.1, considering the classical Hertzian contact problem in two dimensions under uniform mesh refinement. The spatial convergence of the proposed piecewise linear interpolation scheme for quadratic finite elements is also analysed. The improvement in computational complexity is measured in Sect. 8.2 by analysing the three-dimensional contact problem of a half torus going against a complex rigid boundary. Lastly, the proposed piecewise linear interpolation scheme for quadratic finite elements in three dimensions is employed in Sect. 8.3 to solve the contact of a deformable base against a rigid punch.

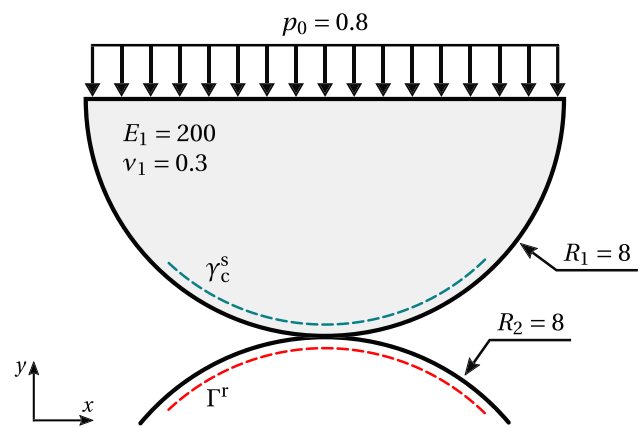

**Fig. 6** Hertzian contact—schematic representation of the problem setting

## 8.1 Spatial convergence—Hertzian contact

The first numerical example to be analysed regards the Hertzian frictionless contact between two cylinders under plane strain conditions. The objective is to assess the spatial convergence rate of the original dual mortar formulation and the proposed techniques. The geometry of the problem is schematically represented in Fig. 6. It consists of one deformable cylinder with radius $R_1 = 8$, described by the St.-Venant-Kirchhoff hyperelastic material model, with Young's modulus $E = 200$ and Poisson's ratio $\nu = 0.3$, contacting against a rigid cylinder with the same radius $R_2 = 8$. For simplicity, only half of the deformable cylinder is modelled and the constant pressure of $p_0 = -0.8$ is applied to the top surface of the hemisphere. The problem is discretised using a structured mesh of standard 4- and 8-noded quadrilateral elements for first- and second-order interpolation. The external pressure is applied incrementally in 20 steps, considering a relative convergence tolerance for the nonlinear solver of $\varepsilon_r = 1 \times 10^{-10}$. Only the segment-based strategy is considered in order to not compromise the spatial convergence properties of the underlying mixed finite element formulation.

Firstly, in order to inspect the convergence rate, successive uniform mesh refinement steps are employed in a structured manner. Figure 7 shows the entire coarse finite element mesh, including a snippet of the remaining refinement levels. The contact region starts with a discretisation based on $n = 4$ elements, which is then doubled four times, i.e., until reaching $n = 64$ elements. The idea is to compare the solution of each level with a reference solution, here obtained using the second-order finite element mesh with $n = 256$ (two refinement levels above the most refined mesh).

Figure 8 shows the discretisation error based on the $H^1$-norm of the error in the displacement field, i.e., by evaluating directly $\|\boldsymbol{u} - \boldsymbol{u}^{\mathrm{h}}\|$ between solutions. For first-order interpolation, $O(h)$ convergence is observed, while for second-order interpolation, optimal results of $O(h^{3/2})$ are achieved. These are in accordance with the theoretical estimates and numerical investigations carried out within the context of unilateral

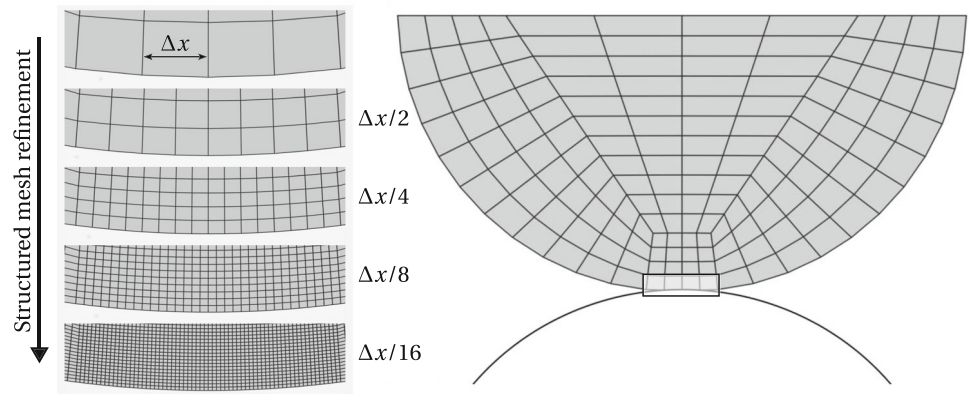

**Fig. 7** Hertzian contact—successive finite element mesh refinement

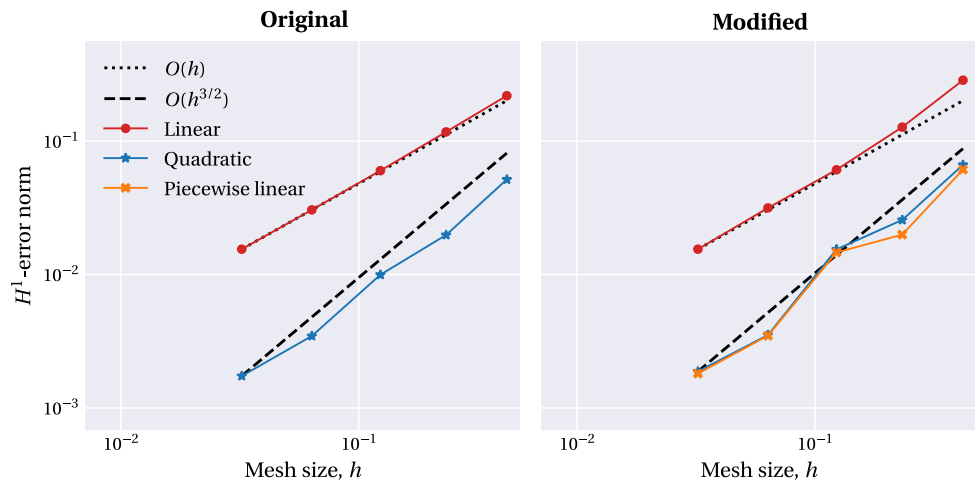

**Fig. 8** Hertzian contact—convergence of the $H^1$-norm of discretization error $\|u - u^h\|$, for both first- and second-order finite interpolation based on quadrilaterals

contact due to a reduced regularity of contact solutions, e.g., see [26,31]. Regarding the ongoing formulation for rigid/deformable contact, one significant result is that the convergence rates of the proposed algorithm are practically identical with the original formulation. This indicates that the mathematical structure of the dual mortar formulation remains unaffected by the employed mixed interpolation scheme. The same is observed for the piecewise linear interpolation.

Numerical results for the contact pressure distribution are illustrated in Fig. 9 for first-order interpolation using different discretization levels.[3] As expected, no oscillations on the contact normal pressure are observed throughout the active contact region, and the solution with the second level of refinement, $n = 8$, is practically identical with the remaining finer meshes. The coarse mesh, $n = 4$, appears to be inappropriate to model the contact problem accurately, as the active region is only described by three elements. Notwithstanding, this problem configuration has been intentionally considered beforehand to identify possible limitations of the formulations. Even in this case, no particular formulation becomes clearly more compelling in terms of accuracy and, therefore, reinforces the conclusion that the porposed formulation behaves similarly to the original.

Figure 10 shows the maximum normal contact pressure, $p_{\max}$, with mesh refinement for all the combinations considered in this section. All the results converge to the same value, with the only significant difference being the results obtained with the coarse mesh. While the first-order the original formulation tends to underestimate the maximum contact pressure, the first-order modified technique tends to follow the trend of second-order interpolation and overestimates the result. The results obtained with the piecewise linear interpolation have the slightest variation.

Lastly, in order to have a complete picture of the results, Fig. 11 shows the deformed configuration of the cylinder, including a coloured representation of the vertical displacement. Only the case with the coarse mesh is represented, as the remaining meshes are similar. In fact, even for the coarse mesh, the difference in the displacement field is so slight that the only significant difference regards the results obtained with the new approach with first-order elements. A small penetration of the nodes at the end of the active contact zone is observed, which reflects the difference in the way the weighted gap is evaluated. Nonetheless, this is consistent with similar studies, e.g., in [25,28], and is a behaviour that tends to vanish with mesh refinement. No overlap is observed in the results obtained with the next finer mesh, $n = 8$.

---

[3] Even though not documented here for the sake of brevity, similar results are obtained with the second-order interpolation.

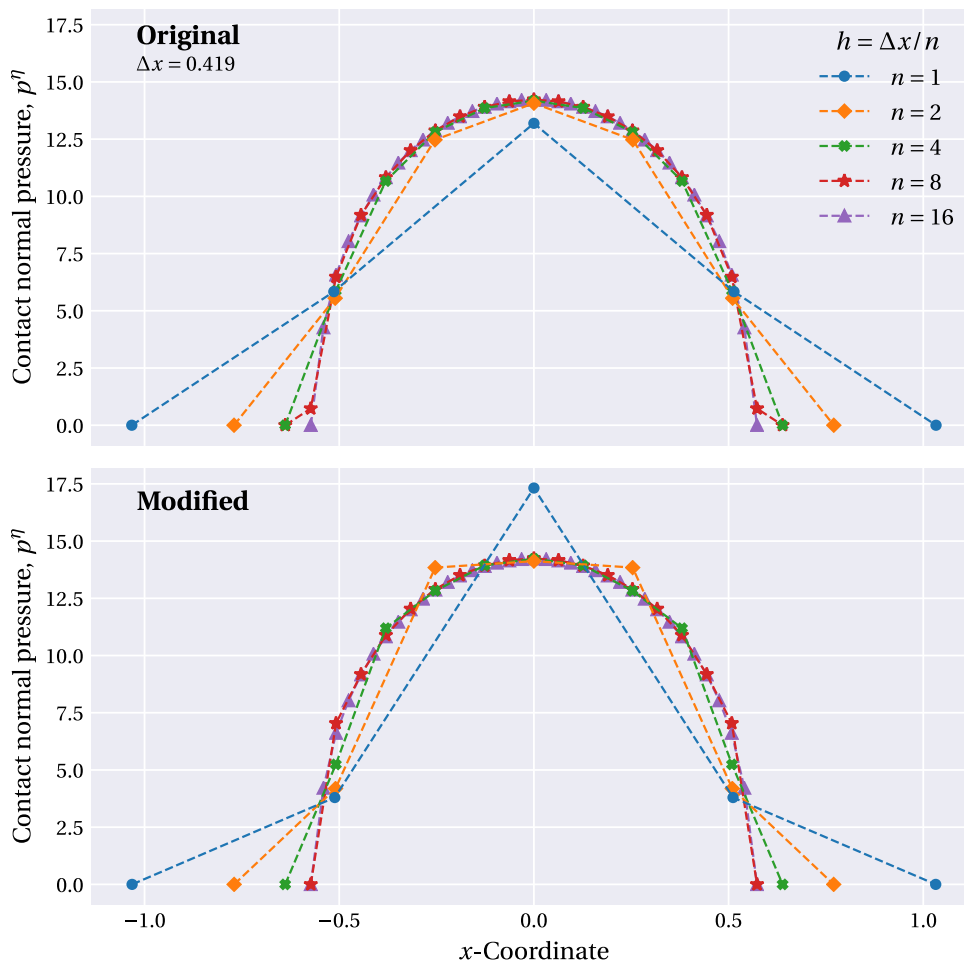

**Fig. 9** Hertzian contact—convergence of the contact pressure distribution $p$ with mesh refinement for first-order quadrilateral meshes

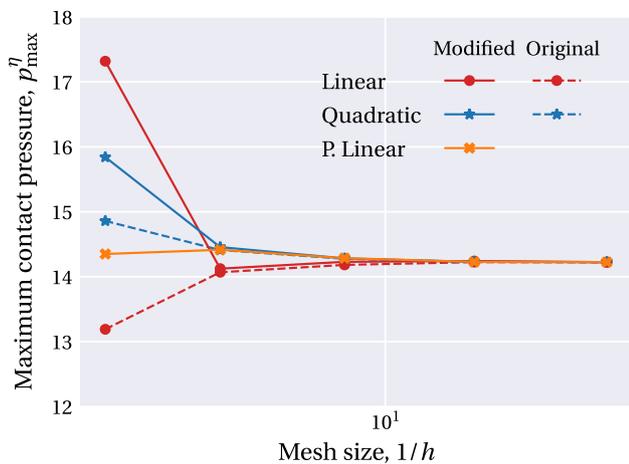

**Fig. 10** Hertzian contact—maximum normal contact pressure $p_{max}$ with mesh refinement for first-order and second-order quadrilateral meshes

### 8.2 Computational complexity—Half-torus Signorini contact

In this section, the computational complexity of dual mortar formulations for rigid/deformable contact is quantitatively measured. Besides robustness, the motivation for the proposed techniques is mainly driven by improving the numerical efficiency of contact modelling. Therefore, a numerical example designed to unveil the computational performance of the contact algorithm has been carried out. The problem setup and dimensions are schematically illustrated in Fig. 12, which consists of the contact of a deformable half-torus against a rigid surface with a relatively complex shape. This example renders a high ratio between the contact interface and total degrees of freedom, while involving a non-trivial rigid contour to define the projected normal vectors. The Neo-Hookean material model is considered for the half-torus, which is subjected to an incremental vertical displacement and has its outer surface set as non-mortar. Frictionless contact is assumed based on four combinations of formulations:

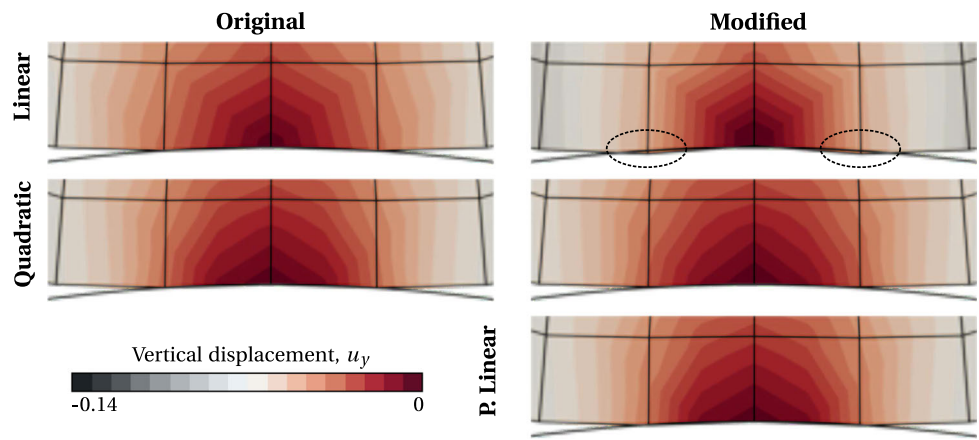

**Fig. 11** Hertzian contact—vertical displacement for different methods. The penetration obtained with the modified weighted gap function and linear finite elements is highlighted

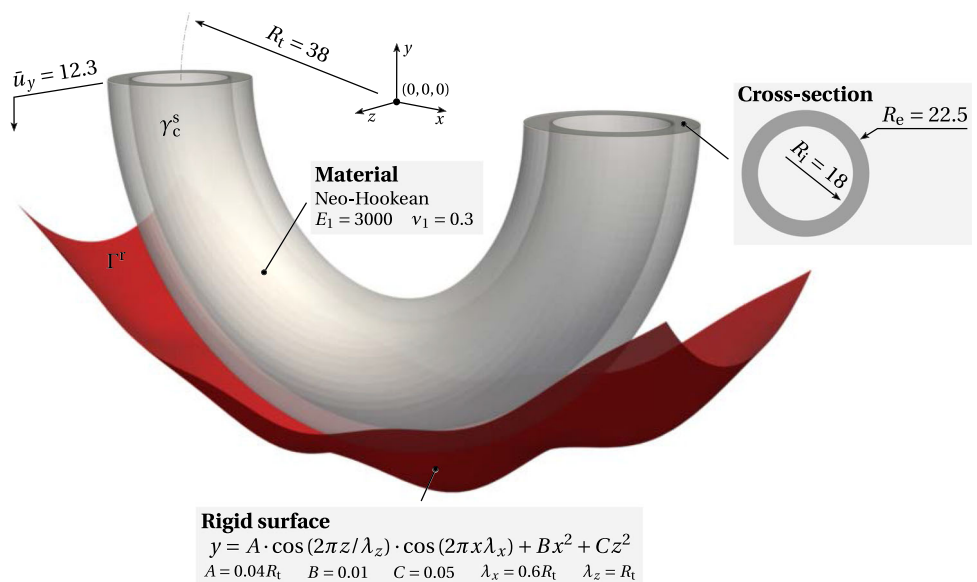

**Fig. 12** Half-torus—schematic representation of the problem setting

- Original/modified algorithm;
- Averaged/projected orthonormal moving frame.

The idea is to measure the impact of the two main aspects discussed in this work (individually and combined), thus getting an estimate of the overall computational performance of the algorithms for rigid/deformable contact.

The body is discretised using a structured 8-noded hexahedron mesh with F-bar elements [34] and the rigid side is discretised using 4-noded bilinear quadrilaterals. The total dimension of the problem is around 85k nodes, from which 13k are rigid. The displacement is applied in 108 equally spaced increments and a relative convergence tolerance of $\varepsilon_r = 1 \times 10^{-6}$ is considered. In order to accentuate the complexity of the contact algorithm, the mortar integrals are evaluated using the segmentation method exclusively. Using a desktop workstation, the total simulation time has varied between 14 and 8 hours. An exemplary representation of the deformed configuration at the end of the simulation is given in Fig. 13, including glyphs representing the contact stress field.

The time needed to evaluate all the terms related to contact is represented in Fig. 14. Here, the average time per iteration of a given increment is plotted against the ratio between the total number of active nodes and the total number of nodes of the finite element mesh. As the number of active nodes increases monotonically with the pseudo-time, the horizontal axis can be interpreted as the pseudo-time, or currently prescribed displacement, yet adjusted to a more relevant quantity for the current analysis. In turn, the ratio of active node numbers measures the impact of contact modelling within the global finite element problem. On the vertical axis, the contact time (mainly dominated by the linearisation update procedures) is summed per iteration and, after the global Newton algorithm converges, it is averaged over the total number of iterations needed to achieve equilibrium conditions. It should be mentioned that, in order evaluate the computational time based on this measure, the number of iterations needed to achieve convergence should be practically the same. Otherwise, the averaging procedure could be neglecting any convergence discrepancies between the methods. In Table 2 the average number of iterations is shown for all the methods. The first observation is that choosing between the averaged or projected normal does not affect the convergence of the algorithm. Modifying the weighted gap, however, increases the average number of iterations by 5%. From a practical perspective, this means that the algorithm will occasionally require one more iteration. Notwithstanding, this also means that one can look at the time per iteration in order to evaluate the computational performance.

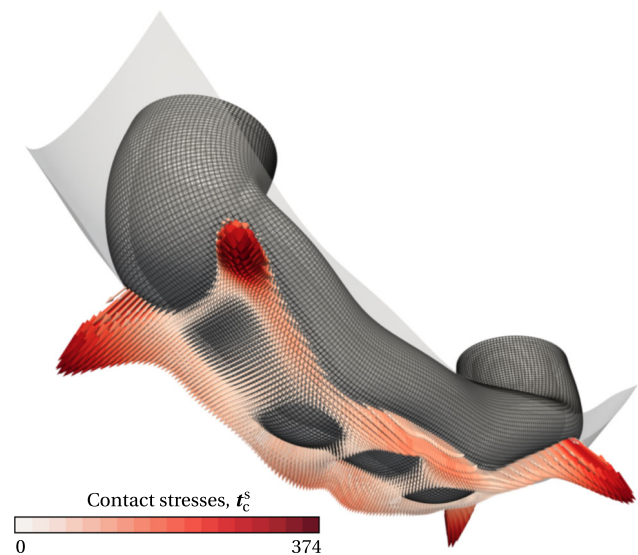

**Fig. 13** Half-torus—deformed configuration contact stress field

It is observed that, as expected, the computational time increases with the total number of active nodes. The classical formulation with the averaged orthonormal frame is the most computational demanding combination. In contrast, the newly proposed methodology based on the modified weighted gap function and projected frame is the fastest. In order to evaluate the differences more clearly, the bottom graph in Fig. 14 shows the speed-up of each formulation in relation to the original method. The most significant improvement in efficiency is achieved by changing the interpolation of the weighted gap, with an average reduction of $\approx 35\%$ in computation time. This reduction is most likely related with avoiding the evaluation of the dual shape functions. The projected frame achieves a reduction of $\approx 10\%$, which means that the two methodologies combined sum up to a $\approx 45\%$ reduction in the computation time for contact evaluation.

At this stage, it seems appropriate to discuss the impact of the contact algorithm within the global framework of finite element modelling. One can acknowledge that the impact of this reduction in the global simulation time is highly dependent on the problem size, the computational implementation and the computer hardware. All the examples presented in this work have been solved using a desktop workstation. The global linear system of equations have been solved by employing a direct solver and all the operations have been carried out in serial. Under these conditions, the problem is dominated by the contact algorithm and, therefore, the global computation times follow approximately the same tendency of the times plotted in Fig. 14. By employing strategies such as parallelisation, one should expect the problem to become dominated more by the linear solver. Nonetheless, even for such an optimised scenario, the computational complexity associated with the contact algorithm (especially the geometrical operations and subsequent linearisation) is typically not negligible. It is expected to remain an essential part of the overall computational performance for more demanding problems with a high ratio of active contact nodes.

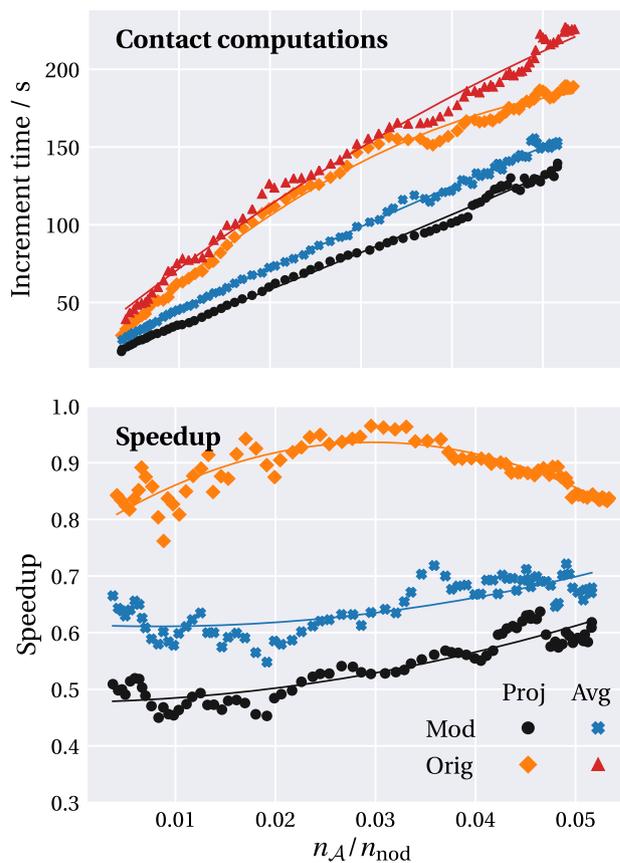

**Fig. 14** Half-torus—time required to complete all operations related with contact (the solid lines are linear fittings)

In order to understand more clearly the reduction in computational complexity, one can look at memory usage. Figure 15 shows the total number of non-zero entries in the derivatives of the unit normal vector at each active contact node for both the averaged and projected techniques. For the sake of simplicity, only the results obtained with the modified weighted gap are shown, as the results of the classical formulation are exactly the same. As expected, this number increases throughout the simulation, and the bandwidth of the derivative of the projected frame is smaller than for the averaged method. The derivative of the unit normal vector appears in every term of the mortar formulation. As already mentioned, it has an amplified effect by dictating the total number of individual operations within the sparse matrix procedures.

**Table 2** Average number of iterations needed to achieve convergence with all the combinations of original and modified weighted gap and averaged and projected unit normal

|  | Averaged | Projected |
| --- | --- | --- |
| Original | 3.7 | 3.7 |
| Modified | 3.9 | 3.9 |

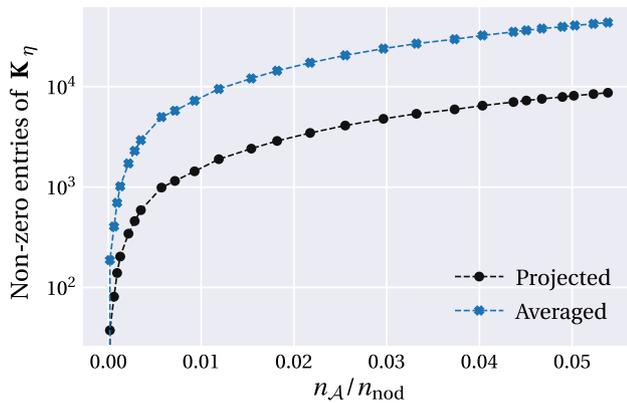

**Fig. 15** Half-torus—number of non-zero entries of the derivative of the unit normal vector

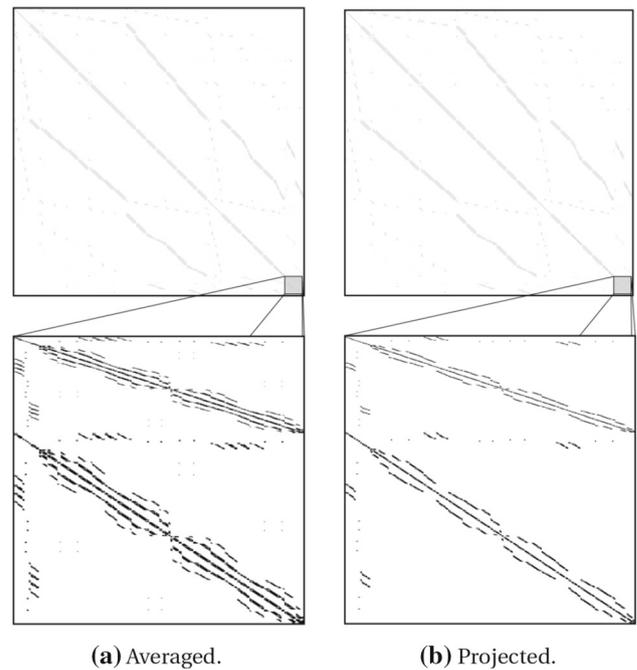

**Fig. 16** Half-torus—exemplary sparsity pattern of the global stiffness matrix, where black pixels represent non-zero entries

Besides the total number of non-zero entries, one can look at the sparsity pattern of the global stiffness matrix to understand the improvements in computational complexity. Figure 16 shows a visualisation of an exemplary square system matrix originating from the averaged and projected methodologies, in which a non-zero entry in the matrix is marked with a black pixel. For ease of interpretation, a magnified representation of the blocks associated with the contact constraints is also included, as any modification in the contact algorithm will be reflected there. As expected, the pattern is more compact in the projected frame variant because the projected frame derivative involves fewer terms than the averaged strategy. Nonetheless, both global system matrices exhibit the pronounced band structure obtained with mortar methods, see e.g., [35].

### 8.3 Piecewise linear interpolation—3D punch

The last numerical example is focused on demonstrating the effect of the piecewise linear interpolation for quadratic dual mortar methods in three dimensions. The problem is schematically represented in Fig. 17 and consists of a deformable cuboid base being pressed against a rigid pin with rounded edges under frictionless conditions. The small fillet radius on the rigid punch naturally produces high local curvatures, especially for coarse discretisation, thus posing substantial convergence problems for algorithms that do not rely on strictly positive interpolation functions for the weighted gap calculation. The base material is characterised by the Neo-Hookean constitutive model. Both the bottom and lateral faces have their vertical displacement prescribed and are fixed along the remaining directions, such that no lateral movement is allowed (thus avoiding unstable configurations). The displacement is applied in 65 equally spaced increments, considering a convergence tolerance of $\varepsilon_r = 1 \times 10^{-6}$.

The deformable base is discretised using 20-noded hexahedra with full integration, while the rigid pin is discretised with 8-noded quadrilateral elements. For the sake of integration accuracy, the mortar integrals are being evaluated using the segmentation method exclusively. The deformed configuration of the base at the end of the simulation is represented in Fig. 18, including a contour plot of the displacement field along the vertical direction. The proposed algorithm converges without any spurious contact states or oscillations for both coarse and fine finite element meshes.

In order to visualise more clearly the concepts of the projected frame and piecewise linear interpolation, a snippet of the integration cells at the final increment of the simulation is schematically represented in Fig. 19. The unit normal vec-tor associated with each non-mortar node is also indicated. Looking at the unit normals, one can see that they all point towards the rigid punch, which does not match the contour of the body for inactive regions of the contact interface (e.g. top left part of the left image in Fig. 19). Nonetheless, as the distance between both surfaces decreases, they become almost identical and the normal vectors start capturing the contour of the non-mortar boundary very precisely. This aspect of the algorithm is also visualised when looking at the orientation of the integration cells, which are established based on the auxiliary plane defined from the continuous field of normal vectors. Lastly, in Fig. 19 the division of the elements into sub-elements can also be identified when looking at the contour of the integration cells.

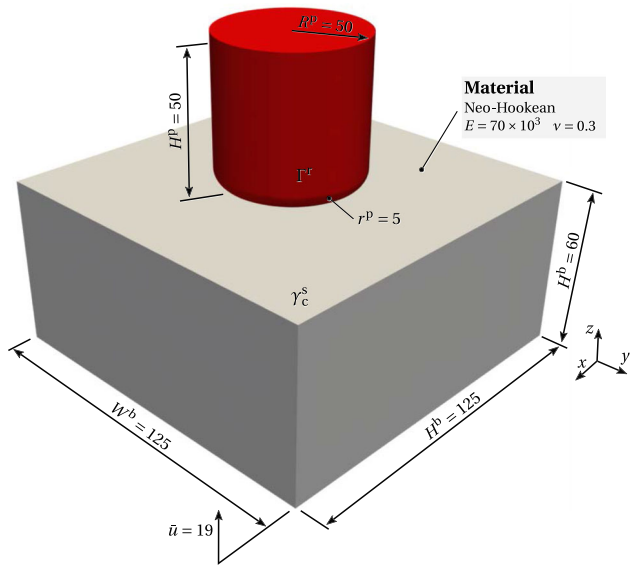

**Fig. 17** 3D punch—schematic representation of the problem setting

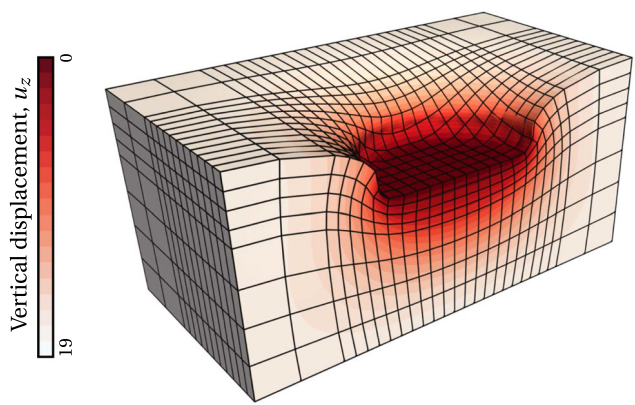

**Fig. 18** 3D punch—vertical displacement of the deformable base at the end of the simulation. Only half of the base is visualized

## 9 Conclusions

This work focuses on the development of an efficient dual mortar contact algorithm specifically tailored for rigid/deformable contact under large deformations. This class of contact problems, also commonly termed as Signorini contact, is found in a wide range of engineering systems. The consideration that one of the contacting bodies is a rigid obstacle unlocks significant simplification potential for computational contact analysis. Moreover, this is a problem configuration commonly found in contact homogenisation as well. It can be traced back to a well-known result of contact mechanics, which states that, under certain circumstances, the contact between two rough surfaces can me mapped to an equivalent roughness and a rigid flat.

The primary motivation for this contribution is to reduce the considerable computational complexity of dual mortar methods, especially in three dimensions. The FEM itself is already regarded as an expensive simulation method, which becomes an even more computationally demanding option when used in combination with mortar methods. Therefore, the simplifications associated with Signorini contact are exploited in order to simplify the algorithm while preserving its accuracy and flexibility—the main argument favouring the mortar FEM. The first idea regards the consideration of standard shape functions to define the weighted gap function instead of dual shape functions. In the particular case of Signorini contact, this methodology is exceptionally advantageous, as it eliminates the need to explicitly evaluate the dual basis during the simulation. The extension of this formulation to second-order interpolation in three dimensions is also carried out by proposing a piecewise linear interpolation for the variation of the Lagrange multipliers and the resulting weighted gap definition. The second idea is based on a new definition for the nodal orthonormal moving frame attached to each contact node, which uses the projection of the frame on the rigid side to the deformable body. When compared with the well-established method based on the averaged unit normal, this technique reduces the bandwidth of the derivatives associated with both the normal and tangential vectors, which now depend only on the degrees of freedom associated with the finite element node itself. This effect propagates throughout all the contact-related computations and, in the end, a significant reduction in the computation time of all the operations related to the mortar approach is achieved.

The numerical examples indicate that, firstly, the algorithm based on the modified weighted gap function preserves the optimal convergence rate of the discretisation error as known from the original formulation. Moreover, the reduction in computational complexity is quantitatively measured, and a reduction of $\approx 40\%$ in the computation time needed to enforce the contact constraints is achieved. Lastly, the proposed piecewise linear interpolation for 3D second-order

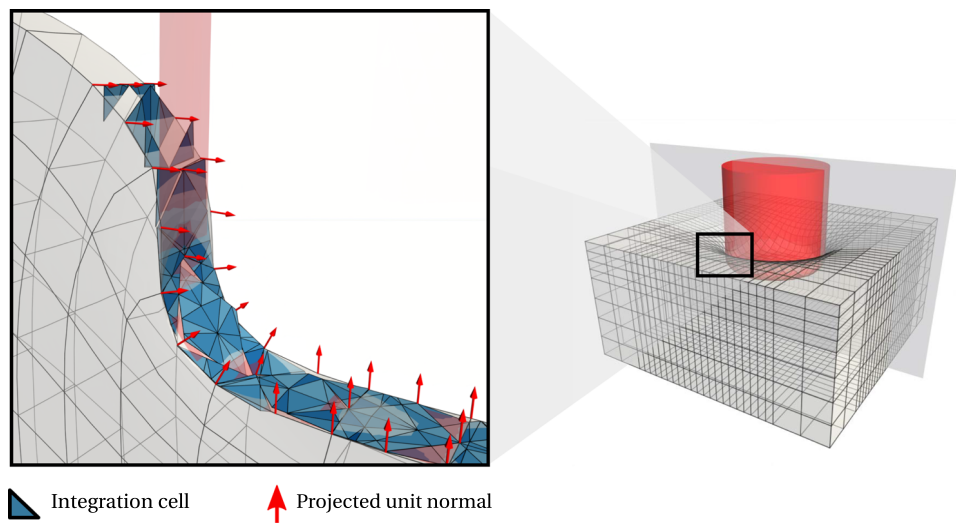

**Fig. 19** 3D punch—slice of the central section of the deformed configuration. The arrows represent the unit normal vector at each non-mortar node and the blue cells are the integration cells resulting from the segmentation algorithm

mortar FEM is applied to a problem with high local curvatures, which typically hinders the application of standard versions of the dual mortar method. This confirms the robustness of the proposed algorithm.

**Funding** Open Access funding enabled and organized by Projekt DEAL.

## Appendix A: Mapping operations between quadratic finite elements and linear sub-elements

The application of the piecewise linear interpolation requires the establishment of proper mappings, herein denoted by $\Lambda$, from the parent element space, $\boldsymbol{\xi}$, to the sub-element space, $\boldsymbol{\xi}^{\text{sub}}$, viz.

$$\boldsymbol{\xi}^{\text{sub}} = \Lambda(\boldsymbol{\xi}). \tag{A.1}$$

Conversely, the inverse mapping from the sub-element space $\boldsymbol{\xi}^{\text{sub}}$ back to the parent element reads

$$\boldsymbol{\xi} = \Lambda^{-1}(\boldsymbol{\xi}^{\text{sub}}). \tag{A.2}$$

These mappings can be derived by employing simple linear transformations between both domains based on geometrical considerations. Table 3 shows the transformations associated with quadratic 6-noded triangles, followed in Table 4 and Table 5 by the 8-noded and 9-noded quadrilaterals, respectively. Besides the expressions for the mappings, linearisation of all deformation-dependent terms in mortar integrals requires taking into account the mapping in the derivative chain rule. Therefore, one also needs to derive the Jacobian matrices associated with the mapping operations, which are represented by the matrices $[\Xi] \in \mathbb{R}^{(d-1)\times(d-1)}$. The construction of these matrices is as follows:

$$[\Xi] \equiv \frac{\partial \boldsymbol{\xi}^{\text{sub}}}{\partial \boldsymbol{\xi}^{\text{s}}} = \begin{bmatrix} \xi^{\text{sub}}_{1,\xi_1} & \xi^{\text{sub}}_{1,\xi_1} \\ \xi^{\text{sub}}_{2,\xi_1} & \xi^{\text{sub}}_{2,\xi_2} \end{bmatrix}. \tag{A.3}$$

Their inverse read

$$[\Xi]^{-1} = \frac{\partial \boldsymbol{\xi}}{\partial \boldsymbol{\xi}^{\text{sub}}} = \begin{bmatrix} \xi_{1,\xi^{\text{sub}}_1} & \xi_{1,\xi^{\text{sub}}_1} \\ \xi_{2,\xi^{\text{sub}}_1} & \xi_{2,\xi^{\text{sub}}_2} \end{bmatrix}. \tag{A.4}$$

**Table 3** Mapping between interface elements and the associated sub-elements for the 6-noded triangle

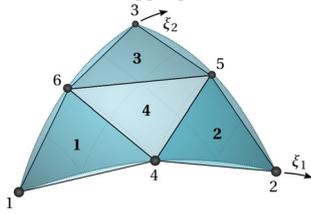

| Sub-element | Node ordering | $\Lambda(\boldsymbol{\xi})$ | $\Xi$ | $\Lambda^{-1}(\boldsymbol{\xi}^{\mathrm{sub}})$ | $\Xi^{-1}$ |
|---|---|---|---|---|---|
| 1 | 1, 4, 6 | $\xi_1^{\mathrm{sub}} = 2\xi_1 \;\; \xi_2^{\mathrm{sub}} = 2\xi_2$ | $\begin{bmatrix} +2 & 0 \\ 0 & +2 \end{bmatrix}$ | $\xi_1 = \xi_1^{\mathrm{sub}}/2 \;\; \xi_2 = \xi_2^{\mathrm{sub}}/2$ | $\begin{bmatrix} +1/2 & 0 \\ 0 & 1/2 \end{bmatrix}$ |
| 2 | 4, 2, 5 | $\xi_1^{\mathrm{sub}} = 2\xi_1 - 1 \;\; \xi_2^{\mathrm{sub}} = 2\xi_2$ | $\begin{bmatrix} +2 & 0 \\ 0 & +2 \end{bmatrix}$ | $\xi = (\xi_1^{\mathrm{sub}} + 1)/2 \;\; \xi_2 = \xi_2^{\mathrm{sub}}/2$ | $\begin{bmatrix} +1/2 & 0 \\ 0 & 1/2 \end{bmatrix}$ |
| 3 | 6, 5, 3 | $\xi_1^{\mathrm{sub}} = 2\xi_1 \;\; \xi_2^{\mathrm{sub}} = 2\xi_2 - 1$ | $\begin{bmatrix} +2 & 0 \\ 0 & +2 \end{bmatrix}$ | $\xi_1 = \xi_1^{\mathrm{sub}}/2 \;\; \xi_2 = (\xi_2^{\mathrm{sub}} + 1)/2$ | $\begin{bmatrix} +1/2 & 0 \\ 0 & +1/2 \end{bmatrix}$ |
| 4 | 5, 6, 4 | $\xi_1^{\mathrm{sub}} = -2\xi_1 + 1 \;\; \xi_2^{\mathrm{sub}} = -2\xi_2 + 1$ | $\begin{bmatrix} -2 & 0 \\ 0 & -2 \end{bmatrix}$ | $\xi_1 = (-\xi_1^{\mathrm{sub}} + 1)/2 \;\; \xi_2 = (-\xi_2^{\mathrm{sub}} + 1)/2$ | $\begin{bmatrix} -1/2 & 0 \\ 0 & -1/2 \end{bmatrix}$ |

**Table 4** Mapping between interface elements and the associated sub-elements for the 8-noded quadrilateral

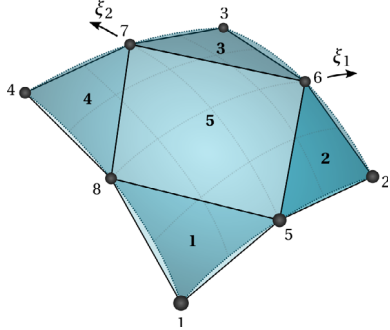

| Sub-element | Node ordering | $\Lambda(\boldsymbol{\xi})$ | $\Xi$ | $\Lambda^{-1}(\boldsymbol{\xi}^{\mathrm{sub}})$ | $\Xi^{-1}$ |
|---|---|---|---|---|---|
| 1 | 1, 5, 8 | $\xi_1^{\mathrm{sub}} = \xi_1 + 1 \;\; \xi_2^{\mathrm{sub}} = \xi_2 + 1$ | $\begin{bmatrix} +1 & 0 \\ 0 & +1 \end{bmatrix}$ | $\xi_1 = \xi_1^{\mathrm{sub}} - 1 \;\; \xi_2 = \xi_2^{\mathrm{sub}} - 1$ | $\begin{bmatrix} +1 & 0 \\ 0 & +1 \end{bmatrix}$ |
| 2 | 2, 6, 5 | $\xi_1^{\mathrm{sub}} = \xi_2 + 1 \;\; \xi_2^{\mathrm{sub}} = -\xi_1 + 1$ | $\begin{bmatrix} 0 & +1 \\ -1 & 0 \end{bmatrix}$ | $\xi_1 = -\xi_2^{\mathrm{sub}} + 1 \;\; \xi_2 = \xi_1^{\mathrm{sub}} - 1$ | $\begin{bmatrix} 0 & -1 \\ +1 & 0 \end{bmatrix}$ |
| 3 | 3, 7, 6 | $\xi_1^{\mathrm{sub}} = -\xi_1 + 1 \;\; \xi_2^{\mathrm{sub}} = -\xi_2 + 1$ | $\begin{bmatrix} -1 & 0 \\ 0 & -1 \end{bmatrix}$ | $\xi_1 = -\xi_1^{\mathrm{sub}} + 1 \;\; \xi_2 = -\xi_2^{\mathrm{sub}} + 1$ | $\begin{bmatrix} -1 & 0 \\ 0 & -1 \end{bmatrix}$ |
| 4 | 4, 8, 7 | $\xi_1^{\mathrm{sub}} = -\xi_2 + 1 \;\; \xi_2^{\mathrm{sub}} = \xi_1 + 1$ | $\begin{bmatrix} 0 & -1 \\ +1 & 0 \end{bmatrix}$ | $\xi_1 = \xi_2^{\mathrm{sub}} - 1 \;\; \xi_2 = -\xi_1^{\mathrm{sub}} + 1$ | $\begin{bmatrix} 0 & +1 \\ -1 & 0 \end{bmatrix}$ |
| 5 | 5, 6, 7, 8 | $\xi_1^{\mathrm{sub}} = \xi_2 + \xi_1 \;\; \xi_1^{\mathrm{sub}} = \xi_2 - \xi_1$ | $\begin{bmatrix} +1 & +1 \\ -1 & +1 \end{bmatrix}$ | $\xi_1 = (\xi_1^{\mathrm{sub}} - \xi_2^{\mathrm{sub}})/2 \;\; \xi_2 = (\xi_1^{\mathrm{sub}} + \xi_2^{\mathrm{sub}})/2$ | $\begin{bmatrix} +1/2 & -1/2 \\ +1/2 & +1/2 \end{bmatrix}$ |

**Table 5** Mapping between interface elements and the associated sub-elements for the 9-noded quadrilateral

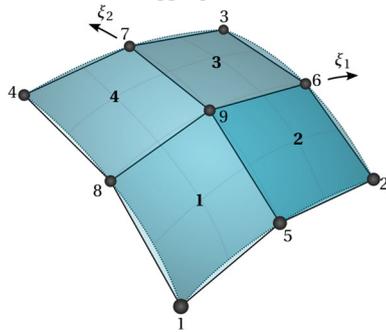

| Sub-element | Node ordering | $\Lambda(\boldsymbol{\xi})$ | $\Xi$ | $\Lambda^{-1}(\boldsymbol{\xi}^{\text{sub}})$ | $\Xi^{-1}$ |
|---|---|---|---|---|---|
| 1 | 1, 5, 9, 8 | $\xi_1^{\text{sub}} = 2\xi_1 + 1 \;\; \xi_2^{\text{sub}} = 2\xi_2 + 1$ | $\begin{bmatrix} +2 & 0 \\ 0 & +2 \end{bmatrix}$ | $\xi_1 = (\xi_1^{\text{sub}} - 1)/2 \;\; \xi_2 = (\xi_2^{\text{sub}} - 1)/2$ | $\begin{bmatrix} +1/2 & 0 \\ 0 & +1/2 \end{bmatrix}$ |
| 2 | 5, 2, 6, 9 | $\xi_1^{\text{sub}} = 2\xi_2 - 1 \;\; \xi_2^{\text{sub}} = 2\xi_1 + 1$ | $\begin{bmatrix} +2 & 0 \\ 0 & +2 \end{bmatrix}$ | $\xi_1 = (\xi_2^{\text{sub}} + 1)/2 \;\; \xi_2 = (\xi_1^{\text{sub}} - 1)/2$ | $\begin{bmatrix} +1/2 & 0 \\ 0 & +1/2 \end{bmatrix}$ |
| 3 | 9, 6, 3, 7 | $\xi_1^{\text{sub}} = 2\xi_1 - 1 \;\; \xi_2^{\text{sub}} = 2\xi_2 - 1$ | $\begin{bmatrix} +2 & 0 \\ 0 & +2 \end{bmatrix}$ | $\xi_1 = (\xi_1^{\text{sub}} + 1)/2 \;\; \xi_2 = (\xi_2^{\text{sub}} + 1)/2$ | $\begin{bmatrix} +1/2 & 0 \\ 0 & +1/2 \end{bmatrix}$ |
| 4 | 8, 9, 7, 4 | $\xi_1^{\text{sub}} = 2\xi_2 + 1 \;\; \xi_1^{\text{sub}} = 2\xi_2 - 1$ | $\begin{bmatrix} +2 & 0 \\ 0 & +2 \end{bmatrix}$ | $\xi_1 = (\xi_1^{\text{sub}} - 1)/2 \;\; \xi_2 = (\xi_2^{\text{sub}} + 1)/2$ | $\begin{bmatrix} +1/2 & 0 \\ 0 & +1/2 \end{bmatrix}$ |